\documentclass[aps,prl,amssymb,superscriptaddress,twocolumn]{revtex4-1}

\usepackage{amsmath,amssymb,braket,fancybox}

\usepackage{graphicx}

\newcommand{\av}[1]{\overline{#1}}
\newcommand{\mc}[1]{\mathcal{#1}}
\newcommand{\mr}[1]{\mathrm{#1}}
\newcommand{\mbb}[1]{\mathbb{#1}}

\newcommand{\lrs}[1]{\left( #1 \right)}
\newcommand{\lrm}[1]{{\left\{ #1 \right\}}}
\newcommand{\lrl}[1]{\left[ #1 \right]}
\newcommand{\lrv}[1]{\left| #1 \right|}

\newcommand{\fracd}[2]{\frac{\mathrm{d} #1 }{\mathrm{d} #2 }}

\newcommand{\aln}[1]{
\begin{align}
#1
\end{align}
}

\newcommand{\ra}{\rightarrow}

\newcommand{\Tr}{\mr{Tr}}

\begin{document}
\title{
Lindbladian Many-Body Localization}
\date{\today}
\author{Ryusuke Hamazaki}
\affiliation{
Nonequilibrium Quantum Statistical Mechanics RIKEN Hakubi Research Team, RIKEN Cluster for Pioneering Research (CPR), RIKEN iTHEMS, Wako, Saitama 351-0198, Japan
}
\author{Masaya Nakagawa}
\affiliation{Department of Physics, University of Tokyo, 7-3-1 Hongo, Bunkyo-ku, Tokyo 113-0033, Japan}
\author{Taiki Haga}
\affiliation{Department of Physics and Electronics, Osaka Metropolitan University, Osaka, 599-8531, Japan}
\author{Masahito Ueda}
\affiliation{Department of Physics, University of Tokyo, 7-3-1 Hongo, Bunkyo-ku, Tokyo 113-0033, Japan}
\affiliation{
RIKEN Center for Emergent Matter Science (CEMS), Wako, Saitama 351-0198, Japan}
\affiliation{
Institute for Physics of Intelligence, University of Tokyo, 7-3-1 Hongo, Bunkyo-ku, Tokyo 113-0033, Japan}

\begin{abstract}

We discover a novel localization transition that alters the  dynamics of coherence in disordered many-body spin systems subject to Markovian dissipation.
The transition occurs in the middle spectrum of the Lindbladian super-operator whose eigenstates obey the universality of non-Hermitian random-matrix theory for weak disorder and exhibit localization of off-diagonal degrees of freedom for strong disorder.
This Lindbladian many-body localization prevents many-body decoherence due to interactions and is conducive to robustness of the coherent dynamics characterized by the rigidity of the decay rate of coherence.

\end{abstract}

\maketitle

\paragraph*{Introduction.}
Strongly disordered potentials of isolated systems significantly alter quantum dynamics.
Spectral statistics of a Hamiltonian characterize thermalizing and non-thermalizing phases.
For weak disorder, the eigenvalue-spacing distribution obeys the Wigner-Dyson statistics, and the eigenstate thermalization hypothesis (ETH)~\cite{Srednicki94,Deutsch91,Rigol08,Rigol09Q,Biroli10,Genway12,Khatami13,Steinigeweg13,Beugeling14,Kim14,Beugeling15,Luitz16,PhysRevLett.120.200604,Hamazaki19R,Khaymovich19,Jansen19,Nation19,Sugimoto21,Fritzsch21,sugimoto2021eigenstate,d2016quantum,mori2018thermalization} holds, reflecting the universality of Hermitian random matrix theory (RMT)~\cite{Haake}.
For strong disorder, eigenvalues obey the Poisson distribution, and eigenstates are described by quasi-local integrals of motion, providing unique features of many-body localization (MBL)~\cite{Basko06,Zunidaric08,Pal10,Gogolin11,Bardarson12,Iyer13,Huse13,Serbyn13,Huse14,Kjall14,Nandkishore14,Bera15,Ponte15,Potter15,Schreiber15,Luitz15,Vosk15,Smith16,Choi16,Imbrie16D,Imbrie16O,Khemani17,alet2018many,lukin2019probing,Abanin19}.

However, no system is immune to external dissipation~\cite{Syassen08,Barreiro11,Yan13,Barontini13,Labouvie15,Patil15,Gao15,Labouvie16,Luschen16,Raftery17,Tomita17,Lapp19,takasu2020pt,bouganne2020anomalous,Kessler21,noel2021observation}, which dramatically alters the nature of MBL~\cite{PhysRevB.90.064203,johri2014numerical,Fischer16,Levy16,Medvedyeva16I,Luschen16,Nieuwenburg17,morningstar2021avalanches,sels2021markovian}.
By coupling  a small bath Hamiltonian to an MBL Hamiltonian, the authors of Refs.~\cite{PhysRevB.90.064203,johri2014numerical} discuss that the signature of MBL on the spectral functions of spins survives under dissipation; however, the total system becomes delocalized and satisfies the ETH. 
Instead, the authors of Refs.~\cite{Fischer16,Levy16,Medvedyeva16I,Luschen16} incorporate dissipation through the Lindblad equation and show that
the local integrals of motion are no longer preserved, rendering the stationary state delocalized.
These works mainly focus on how the signature of the Hermitian MBL is altered by dissipation.
However, a question remains concerning whether a sharp localization transition exists as a \textit{unique} phenomenon for a dissipative disordered many-body system.
For example, is there a spectral transition described by the \textit{Lindbladian super-operator} as in the Hamiltonian operator in the  Hermitian MBL~\cite{prev_foot}, and, if any, what is the consequence of the transition on the dynamics?
Note that the non-Hermitian MBL~\cite{hamazaki19nonhermitian} describes the post-selected dynamics of continuously measured  systems via an effective non-Hermitian Hamiltonian and is inapplicable to generic Lindbladian dynamics.

\begin{figure}
\begin{center}
\includegraphics[width=\linewidth]{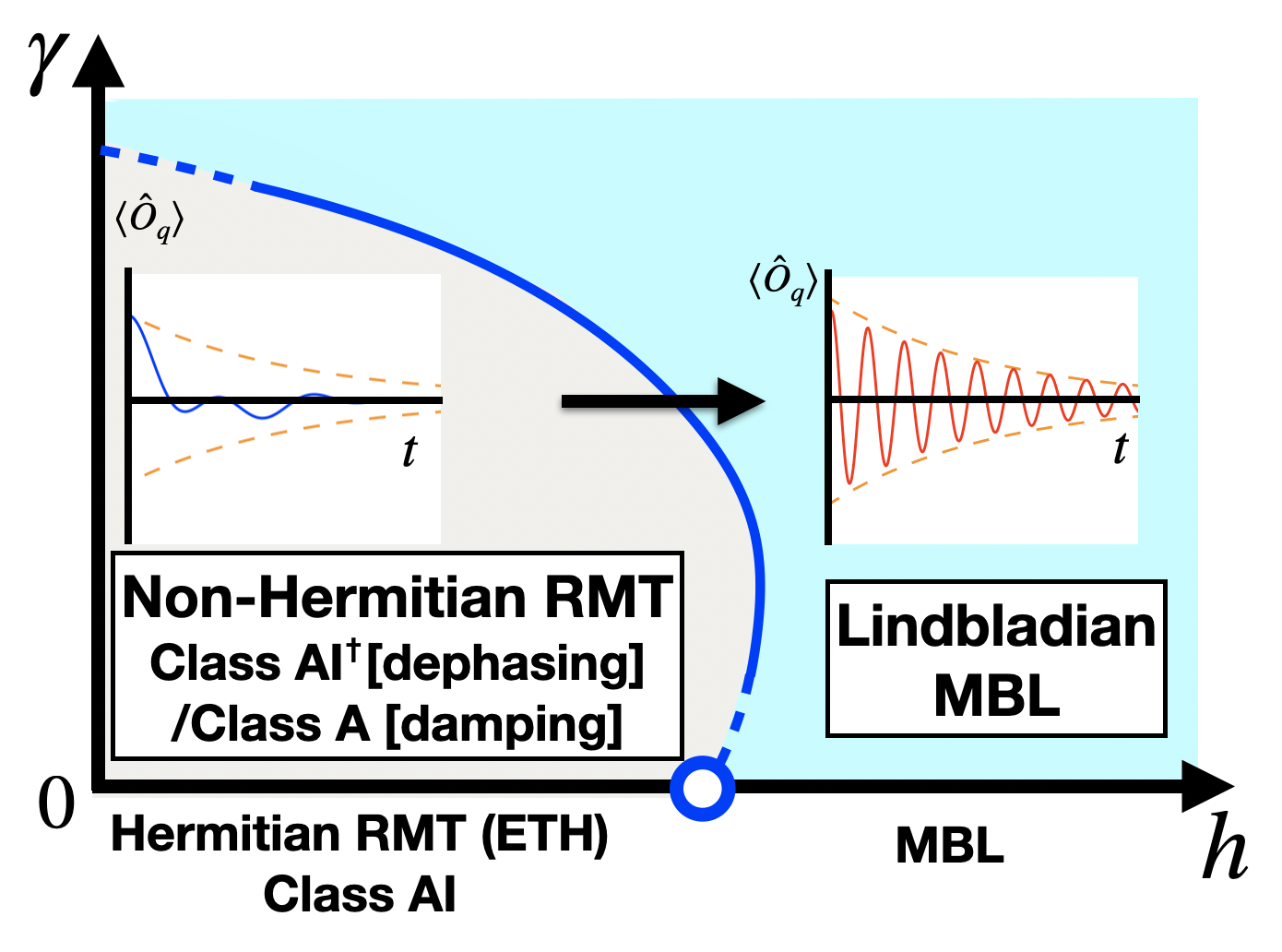}
\end{center}
\caption{
Phase diagram of a dissipative many-body system with disorder.
For finite dissipation $\gamma>0$ and sufficiently strong disorder $h$, the oscillation of an observable $\hat{O}_q$ corresponding to quantum coherence attenuates slowly (red solid curve in the right inset), whereas for weak disorder it decays rapidly (blue solid curve in the left inset). Note that the decay rate is stabilized due to the Lindbladian many-body localization (see the main text).
These two regimes are distinguished by the spectral statistics featuring the localization and non-Hermitian RMT of the Lindbladian super-operator, as a dissipative generalization of the Hermitian Hamiltonian case ($\gamma=0$).
The non-Hermitian RMT classes depend on the type of symmetry of the Lindblad dynamics that is determined by dephasing or damping.
}
\label{intro}
\end{figure}

In this work, we show that an unconventional localization  occurs for dissipative systems as a spectral transition of the {Lindbladian super-operator} (rather than the Hamiltonian), which alters the coherent dynamics.
In fact, spectral statistics in the middle of the spectrum are characterized by the universality of the {non-Hermitian} RMT for weak disorder and by the localization of the  Lindbladian eigenstates  for strong disorder (see Fig.~\ref{intro}). 
We call the latter unique open dissipative localization as the Lindbladian MBL.
Deep in the Lindbladian MBL regime, quantum coherence exhibits a rigid decay whose rate is essentially determined by the  decoherence rate for a single-spin system and stabilized at a roughly constant value against the variations of interaction terms and the magnetic field.
This rigidity of the decay rate is attributed to localization of off-diagonal degrees of freedom in the Lindbladian eigenstates.
The behavior is distinct from interaction-induced many-body decoherence for weak disorder, where eigenmodes with  many  different frequencies and decay rates govern the dynamics.
Using a prototypical Ising model with a dephasing- or damping-type of dissipation, we numerically demonstrate the transition with, e.g., the complex spacing distributions of eigenvalues and the operator-space entanglement entropy (OSEE) of eigenstates.

\paragraph*{Dissipative spin chains with disorder.}
We consider a one-dimensional Ising model with transverse and longitudinal fields under dissipation.
The dynamics is described by the Lindblad equation with $\hbar=1$~\cite{lindblad_1976}:
\aln{
\fracd{\hat{\rho}}{t}=\mc{L}[\hat{\rho}]=-i[\hat{H},\hat{\rho}]+\sum_l\frac{1}{2}\lrl{2\hat{\Gamma}_l\hat{\rho}\hat{\Gamma}_l^\dag-\lrm{\hat{\Gamma}_l^\dag\hat{\Gamma}_l,\hat{\rho}}}.
}
Here, the Hamiltonian is given by
$
\hat{H}=\sum_{i=1}^{L-1}J\hat{\sigma}_i^z\hat{\sigma}_{i+1}^z+\sum_{i=1}^{L}g\hat{\sigma}_i^x+\sum_{i=1}^{L}h_i\hat{\sigma}_i^z,
$
where  $h_i$ is taken randomly from $[-h,+h]$.
Without dissipation, $\hat{H}$ exhibits MBL for sufficiently strong disorder~\cite{Imbrie16D,Imbrie16O}.
We consider two types of dissipation, namely  dephasing
$
\hat{\Gamma}_l=\sqrt{\gamma}\hat{\sigma}^z_l,
$
and damping
$
\hat{\Gamma}_l={\sqrt{\frac{\gamma}{2}}}\hat{\sigma}^-_l
$
with $\hat{\sigma}^-_l=\hat{\sigma}^x_l-i\hat{\sigma}^y_l$.
Below we mainly consider the case with weak dissipation ($\gamma<g$).
The case with large dissipation $(\gamma>g)$ is discussed at  the \textit{Discussion} section.

\begin{figure}
\begin{center}
\includegraphics[width=\linewidth]{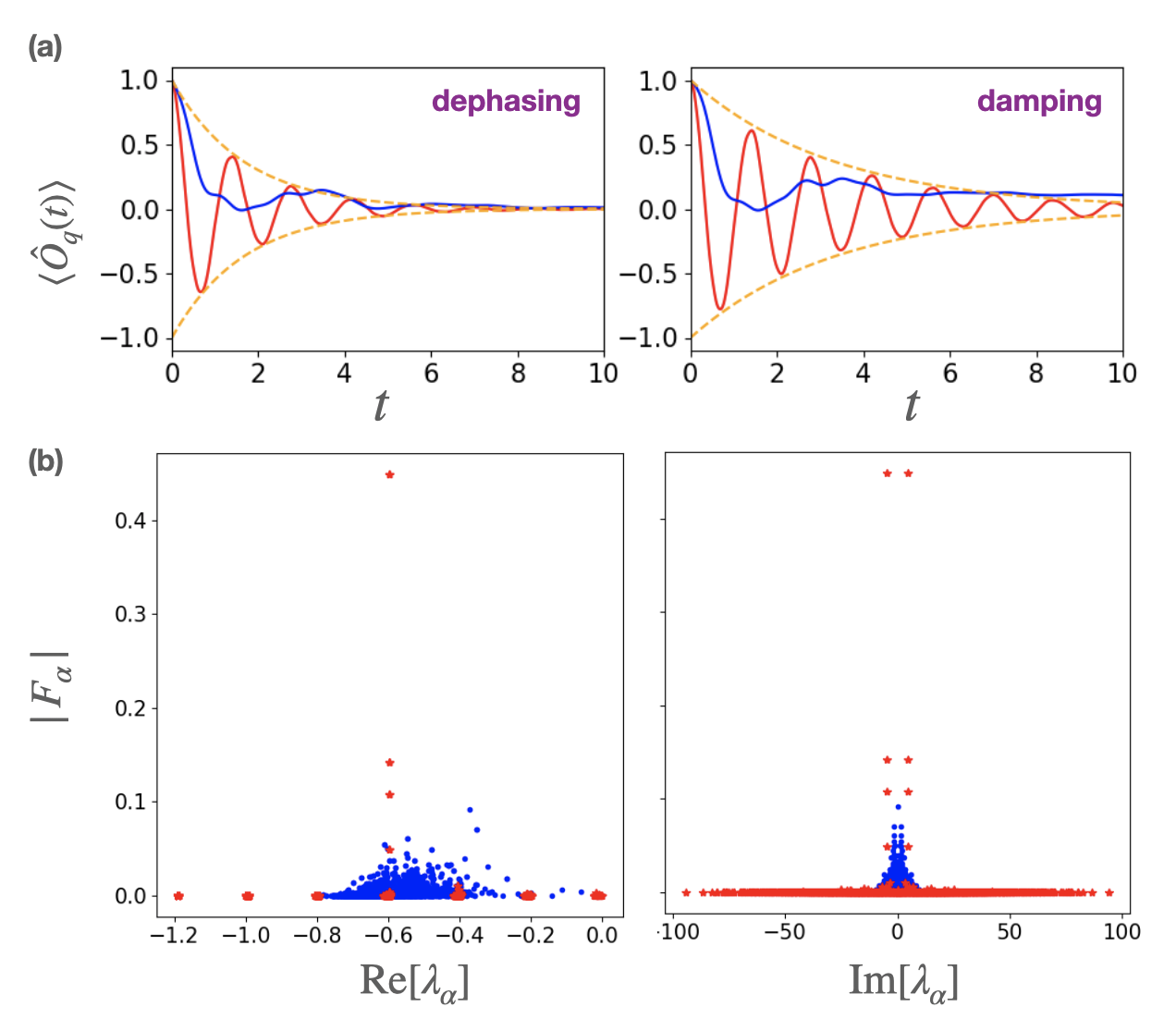}
\end{center}
\caption{
(a) Time evolution of $\braket{\hat{O}_q(t)}$ for $h=1.2$ (blue) and $h=10$ (red).
While $\braket{\hat{O}_q(t)}$  rapidly relaxes to a stationary state for small $h$, it exhibits a slower oscillatory decay  for large $h$.
The orange dashed curves show $\pm e^{-2q\gamma t}$ for the dephasing and $\pm e^{-q\gamma t}$ for the damping type of dissipation, which approximate the amplitudes of the decay for large $h$.
(b) The distribution of $|F_\alpha|$ as a function of the eigenvalue $\lambda_\alpha$ for the dephasing-type dissipation.
While $|F_\alpha|$ in Eq.~\eqref{tevolve} spreads over many $\alpha$'s  for $h=1.2$ (blue dots),  peaks appear for several  eigenstates embedded in the middle of the spectrum, and the other $|F_\alpha|$ are vanishingly small for $h=10$ (red asterisks).
We use  $L=6, J=1, g=-0.9,\gamma =0.1$ and $q=3$.
}
\label{time}
\end{figure}

We first discuss the time evolution of coherence measured by
$
\hat{O}_q=\bigotimes_{k=L-q+1}^{L}\hat{\sigma}_k^x
$
starting from an initial state $\hat{\rho}_q=\ket{\psi_q}\bra{\psi_q}$, where
$
\ket{\psi_q}=(\ket{\downarrow}_1\cdots\ket{\downarrow}_{L-q})\lrs{\frac{\ket{\uparrow}_{L-q+1}\cdots\ket{\uparrow}_L+\ket{\downarrow}_{L-q+1}\cdots\ket{\downarrow}_L}{\sqrt{2}}}.
$
This choice corresponds to measuring the single-spin coherence for $q=1$ and macroscopic coherence of the Greenberger-Horne-Zeilinger state for $q=L$.
In Fig.~\ref{time}(a), we show the time evolution of the quantum  expectation value of $\braket{\hat{O}_q(t)}$ (not averaged over disorder) for different values of disorder strength.
While $\braket{\hat{O}_q(t)}$ rapidly relaxes to a stationary value for small $h$, it exhibits a slower oscillatory decay for large $h$.
Only for the latter case, the decay rate is almost stabilized at around $2\gamma q~(\gamma q)$ for dephasing (damping)~\cite{poly_foot}, which is explained by the decoherence rate for a single-spin system and robust against many-body interactions (see Supplemental Materials~\cite{Supp_foot}).

To understand the distinction between two regimes, we consider the spectral decomposition of the dynamics~\cite{gong2022bounds}:
\aln{\label{tevolve}
\braket{\hat{O}_q(t)}&=\sum_\alpha \frac{\Tr[\hat{O}_q \hat{R}_\alpha]\Tr[ \hat{L}_\alpha^\dag \hat{\rho}_q]}{\Tr[\hat{L}_\alpha^\dag \hat{R}_\alpha]}e^{\lambda_\alpha t}=:\sum_\alpha F_\alpha e^{\lambda_\alpha t},
}
where $\hat{R}_\alpha$ ($\hat{L}_\alpha$) is a right (left) eigenstate of the Lindbladian super-operator $\mc{L}$ with an eigenvalue $\lambda_\alpha\in\mbb{C}$~\cite{normalize_foot}.
For weak disorder, many eigenstates for different $\lambda_\alpha$ contribute to the dynamics, i.e., $|F_\alpha|$ in Eq.~\eqref{tevolve} spreads over many $\alpha$ (see Fig.~\ref{time}(b)), indicating complicated many-body decoherence characterized by a large number of frequencies and decay rates.
On the other hand, for strong disorder, peaks in   $|F_\alpha|$ appear for several  eigenstates in the middle of the spectrum with the other $|F_\alpha|$ being vanishingly small.
These selected eigenstates lead to an oscillatory decay governed by only a few frequencies and a rigid decay rate despite many-body interactions.

\paragraph*{Non-Hermitian random-matrix universality and many-body decoherence.}
The distinctive behavior of the behavior of $|F_\alpha|$ discussed above is attributed to the different regimes characterized by the spectral statistics of the Lindbladian super-operator.
For weak disorder, we find that the statistics obey the non-Hermitian RMT.
In particular, the trace factors appearing in Eq.~\eqref{tevolve} in the middle of the spectrum and away from the real axis are described as
\aln{\label{conj1}
\Tr[\hat{O} \hat{R}_\alpha]\sim \mc{A}_{\hat{O}} (\lambda_\alpha)r_{\alpha},
}
\aln{\label{conj2}
\Tr[\hat{L}_\alpha^\dag\hat{\rho} ]\sim \mc{B}_{\hat{\rho}} (\lambda_\alpha)r_{\alpha},
}
and 
\aln{\label{conj3}
\Tr[\hat{L}_\alpha^\dag \hat{R}_\alpha]^{-1}\sim 
\mc{C}(\lambda_\alpha)r'_{\alpha},
}
where
$\mc{A}_{\hat{O}}$, $\mc{B}_{\hat{\rho}}$, and $\mc{C}$ are smooth functions of $\lambda_\alpha$, and 
$r_\alpha$ and $r'_\alpha$ are complex random variables normalized as $\mbb{E}[|r_\alpha|]=\mbb{E}[|r_\alpha'|]=1$. 
These forms indicate that every energy eigenstate fluctuates randomly within a sufficiently small two-dimensional eigenvalue window where $\mc{A}_{\hat{O}} (\lambda_\alpha)$, $\mc{B}_{\hat{\rho}} (\lambda_\alpha)$, and $\mc{C} (\lambda_\alpha)$ are almost constant.
In this sense, Eqs.~\eqref{conj1}-\eqref{conj3} constitute a  dissipative generalization of the (off-diagonal) ETH for Hermitian systems, which states that $\braket{E_a|\hat{O}|E_b}$ fluctuates  within the small energy windows around $E_a$ and $E_b$ according to (Hermitian) RMT~\cite{Khatami13} for quantum chaotic systems~\cite{classical_foot}.

\begin{figure}
\begin{center}
\includegraphics[width=\linewidth]{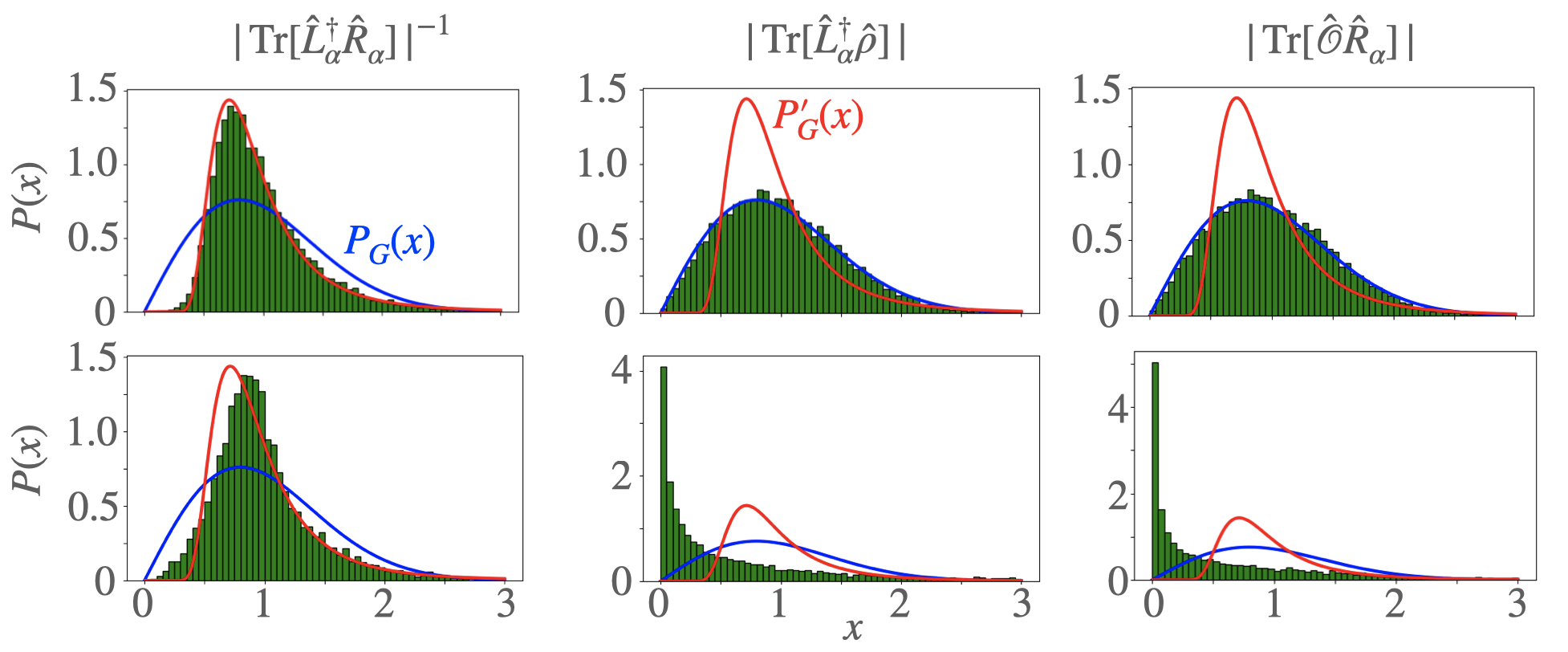}
\end{center}
\caption{Distributions of the normalized values of $|\Tr[ \hat{L}_\alpha^\dag\hat{R}_\alpha]|^{-1}, |\Tr[ \hat{L}_\alpha^\dag\hat{\rho}]|,$ and $|\Tr[\hat{O}\hat{R}_\alpha]|$, i.e., $r_\alpha'$ and $r_\alpha$, for  $h=1.2$ (top) and $h=4.0$ (bottom).
We see that $P(|r'_\alpha|=x)$ and  $P(|r_\alpha|=x)$ obey non-Hermitian random-matrix distributions $P'_G$ and $P_G$, respectively, for sufficiently small $h$.
The distributions deviate from the non-Hermitian random-matrix predictions for large $h$.
We use  $q=1$, $L=7, J=1, g=-0.9$, and $\gamma =0.5$ with  dephasing-type dissipation. 
The distributions are calculated from eigenstates within a small range of eigenvalues with 50 ensembles.
}
\label{dist}
\end{figure}

Notably, the distributions of $r_\alpha$ and $r'_\alpha$ obey the non-Hermitian 
RMT universality~\cite{Ginibre65,Chalker98,mehlig2000statistical}.
Specifically, the distributions $P(r_\alpha)$ and $P(r'_\alpha)$ are described by 
$
P_G(x):=P(|r_\alpha^G|=x)=\frac{\pi}{2}xe^{-\frac{\pi}{2}x^2},
$
which is obtained from the Porter-Thomas distribution, and
$
P_G'(x):=P(|r_\alpha^{G'}|=x)=\frac{32}{\pi^2 x^5}e^{-\frac{4}{\pi x^2}},
$
which is obtained from the results of  a complex Ginibre ensemble~\cite{bourgade2019distribution,realG_foot}.
As shown in Fig.~\ref{dist}, these results are valid for our model with weak disorder, but  invalid for strong disorder~\cite{peak_foot}.

As detailed in Supplemental Material~\cite{Supp_foot},  $|\Tr[\hat{O} \hat{R}_\alpha]|$ and $|\Tr[\hat{L}_\alpha^\dag\hat{\rho}]|$  decrease, and $|\Tr[ \hat{L}_\alpha^\dag\hat{R}_\alpha]|^{-1}$ increases with increasing $L$.
Furthermore,
 $|\Tr[ \hat{L}_\alpha^\dag\hat{R}_\alpha]|^{-1}$ for small $h$ is almost proportional to the dimension $D=2^L$ of the Hilbert space.
We also find $\Tr[\hat{L}_\alpha^\dag\hat{\rho}]\propto D^{-1}$ for small $h$.
These scalings agree with the prediction of the non-Hermitian RMT. 
On the other hand, for strong $h$, the scaling behavior differs from the RMT for both cases.
Note that  the scaling behavior is not simple for $|\Tr[ \hat{O}\hat{R}_\alpha]|$ because of the locality of the operator and that of the Lindbladian.
In this case, we argue that $|\Tr[ \hat{O}\hat{R}_\alpha]|\sim e^{-c'|\lambda_\alpha|}\:(c'>0)$~\cite{Supp_foot}.

The above discussions indicate that, for weak disorder, eigenstates within a small eigenvalue window fluctuate randomly  for  $|\Tr[ \hat{R}_\alpha^\dag\hat{L}_\alpha]|^{-1}, |\Tr[ \hat{L}_\alpha^\dag\hat{\rho}]|,$ and $|\Tr[\hat{O}\hat{R}_\alpha]|$.
Consequently, $F_\alpha$ also behaves randomly without irregular eigenstates as shown in Fig.~\ref{time}(b), resulting in many-body decoherence governed by many modes with different frequencies and decay rates.
The above discussion establishes the previously unknown connection between dissipative quantum chaos and many-body decoherence in terms of eigenstates with the RMT universality.

\paragraph*{Lindbladian MBL.}
We next discuss the strong-disorder case.
We start from the phenomenological picture with quasi-local bits~\cite{Huse14} in the MBL phase without dissipation.
Then, the Hamiltonian reads
 \aln{\label{LIOM}
\hat{H}=\sum_l \tilde{h}_l\hat{\tau}_l^z+\sum_{lm}J_{lm}\hat{\tau}_l^z\hat{\tau}_m^z+
\sum_{lmn}J_{lmn}\hat{\tau}_l^z\hat{\tau}_m^z\hat{\tau}_n^z+\cdots,
}
where $\tilde{h}_l\simeq h_l$, and $J_{lm},J_{lmn},\cdots$ decay exponentially for any far apart two-site indices.
The local integral of motion $\hat{\tau}_l^z$ has a large overlap with $\hat{\sigma}_l^z$.
The Hamiltonian $\hat{H}$ has eigenstates labeled by the eigenvalues $\pm 1$ of $\hat{\tau}_l^z$, i.e., $\ket{\tau_1\cdots\tau_L}$.
We can express  $\hat{\sigma}_l$ by $\hat{\tau}_l$ as
$
\hat{\sigma}_l^z=\sum_{\alpha=x,y,z}Z_l^\alpha\hat{\tau}_l^\alpha+\sum_{jk}\sum_{\alpha,\beta=x,y,z}G_{l,jk}^{\alpha\beta}\hat{\tau}_j^\alpha\hat{\tau}_k^\beta+\text{(higher order terms)}
$,
where $Z_l^z=1-\mr{O}((g/h)^2)$, $Z_l^{x,y}=\mr{O}(g/h)$, and $G_{l,jk}^{\alpha\beta}$ rapidly decays as a function of $|l-j|$ and $|l-k|$.
Also, $G_{l,jk}^{\alpha\beta}=\mr{O}((g/h)^2)$~\cite{Imbrie16D,Imbrie16O}.
A similar representation is obtained for $\hat{\sigma}_l^-$.
Then, the Lindbladian reads $\mc{L}[\hat{\rho}]=-i[\hat{H},\hat{\rho}]+\mc{L}_Z+\mc{L}_P$, where $\mc{L}_Z=\frac{{Z_l^z}^2}{2}\sum_l\lrl{2\hat{\Gamma}_l'\hat{\rho}\hat{\Gamma}_l'^\dag-\lrm{\hat{\Gamma}_l'^\dag\hat{\Gamma}_l',\hat{\rho}}}$ with $\hat{\Gamma}_l'=\sqrt{\gamma}\hat{\tau}_l^z$ (dephasing) or $\hat{\Gamma}_l'=\sqrt{\frac{\gamma}{2}}\hat{\tau}_-^z$ (damping), and $\mc{L}_P$ denotes the remaining perturbation.

We briefly discuss the case of dephasing-type dissipation here (see Supplemental Material for details and the case of damping-type dissipation~\cite{Supp_foot}).
The eigenstates of $-i[\hat{H},\hat{\rho}]+\mc{L}_Z$ can be written as
$\hat{R}_\alpha=\hat{L}_\alpha=\ket{\tau_1\cdots\tau_L}\bra{\tau_1'\cdots\tau_L'}=\hat{\phi}_1^{\alpha_1}\otimes\hat{\phi}_2^{\alpha_2}\otimes\cdots\otimes\hat{\phi}_L^{\alpha_L}$.
Here, $\otimes$ represents the tensor product for different localized bits $\hat{\tau}_{i}$, and $\hat{\phi}_l^{\alpha_l}=\ket{\tau_l}\bra{\tau_l'}$.
Specifically, we take $\hat{\phi}^1=(\hat{\phi}^2)^\dag=\ket{+1}\bra{-1}$, $\hat{\phi}^3=\ket{+1}\bra{+1}$, and 
$\hat{\phi}^4=\ket{-1}\bra{-1}$, where $\ket{\pm 1}$ is the eigenstate of $\hat{\tau}^z$ with an eigenvalue $\pm 1$.
The corresponding eigenvalues are $-i\sum_l2\tilde{h}_l(\delta_{\alpha_l,1}-\delta_{\alpha_l,2})-i(\text{small terms including } J_{lm}, \cdots)-\sum_l2{Z_l^z}^2\gamma(\delta_{\alpha_l,1}+\delta_{\alpha_l,2})$.

When we add $\mc{L}_P$, $\hat{\phi}_l^3$ and
$\hat{\phi}_l^4$ in different eigenstates are mixed, since the eigenvalue difference is almost zero in this case.
In contrast,  $\hat{\phi}_l^1$ and
$\hat{\phi}_l^2$ are typically  stable under first-order perturbation, since the transition matrix elements over the  eigenvalue difference are $\mr{O}(\gamma g/h^2)\ll 1$~\cite{Supp_foot}.
It follows then that while 
diagonal degrees of freedom (DDOF) $\hat{\phi}_l^3$ and
$\hat{\phi}_l^4$ can delocalize, off-diagonal degrees of freedom (ODDOF) $\hat{\phi}_l^1$ and
$\hat{\phi}_l^2$ can localize.
Consequently, eigenstates are not fully mixed, and the  RMT prediction breaks down.
In the dynamics of coherence considered in Fig.~\ref{time}~\cite{Coh_foot}, only the modes with  $\alpha_{L-q+1}=\cdots =\alpha_L=1$ or 2 contribute to $F_\alpha$, and the other modes make negligible contributions to $F_\alpha$.
The  decay rate is then stabilized, and the system evades many-body decoherence.
In particular, the decay rate becomes $\sim 2q\gamma (1-\mr{O}((g/h)^2))$.
For the damping case, the decay rate is similarly stabilized at $\sim q\gamma (1-\mr{O}((g/h)^2))$.

The stabilized dynamics of the coherence (transverse relaxation)  is understood from the nontrivial Lindbladian localization of the ODDOF, which cannot be captured by the classical effective rate equation~\cite{Fischer16,Medvedyeva16I} used for describing the slow longitudinal relaxation in previous studies.
Indeed, the DDOF that govern the longitudinal relaxation~\cite{Fischer16,Medvedyeva16I} are delocalized.
A detailed discussion of the difference between longitudinal and transverse relaxation is given in Supplemental Material~\cite{Supp_foot}.

A few remarks are in order here.
First, the delocalization/localization of DDOF/ODDOF is reasonable because they undergo zero/strong random fields $\sim 0/\pm 2\tilde{h}_l$ in the matrix representation of the Lindbladian~\cite{Supp_foot}.
Second, the localization of ODDOF indicates the emergence of a quasi-local weak symmetry~\cite{buvca2012note,PhysRevA.89.022118}  $\hat{U}_w$ of the Lindbladian, i.e., $\hat{U}_w\mc{L}[\hat{\rho}]\hat{U}_w^\dag=\mc{L}[\hat{U}_w\hat{\rho}\hat{U}_w^\dag]$, which block-diagonalizes the Lindbladian.
Third, as for the discussion of the Hermitian MBL, it is not easy to show  the existence of localization and the precise transition point when we consider higher-order perturbations and resonant states~\cite{vsuntajs2020quantum,morningstar2021avalanches,sels2021markovian}.
We leave it as a future problem to investigate larger system sizes.

\paragraph*{Delocalization-localization transition.}

\begin{figure}
\begin{center}
\includegraphics[width=\linewidth]{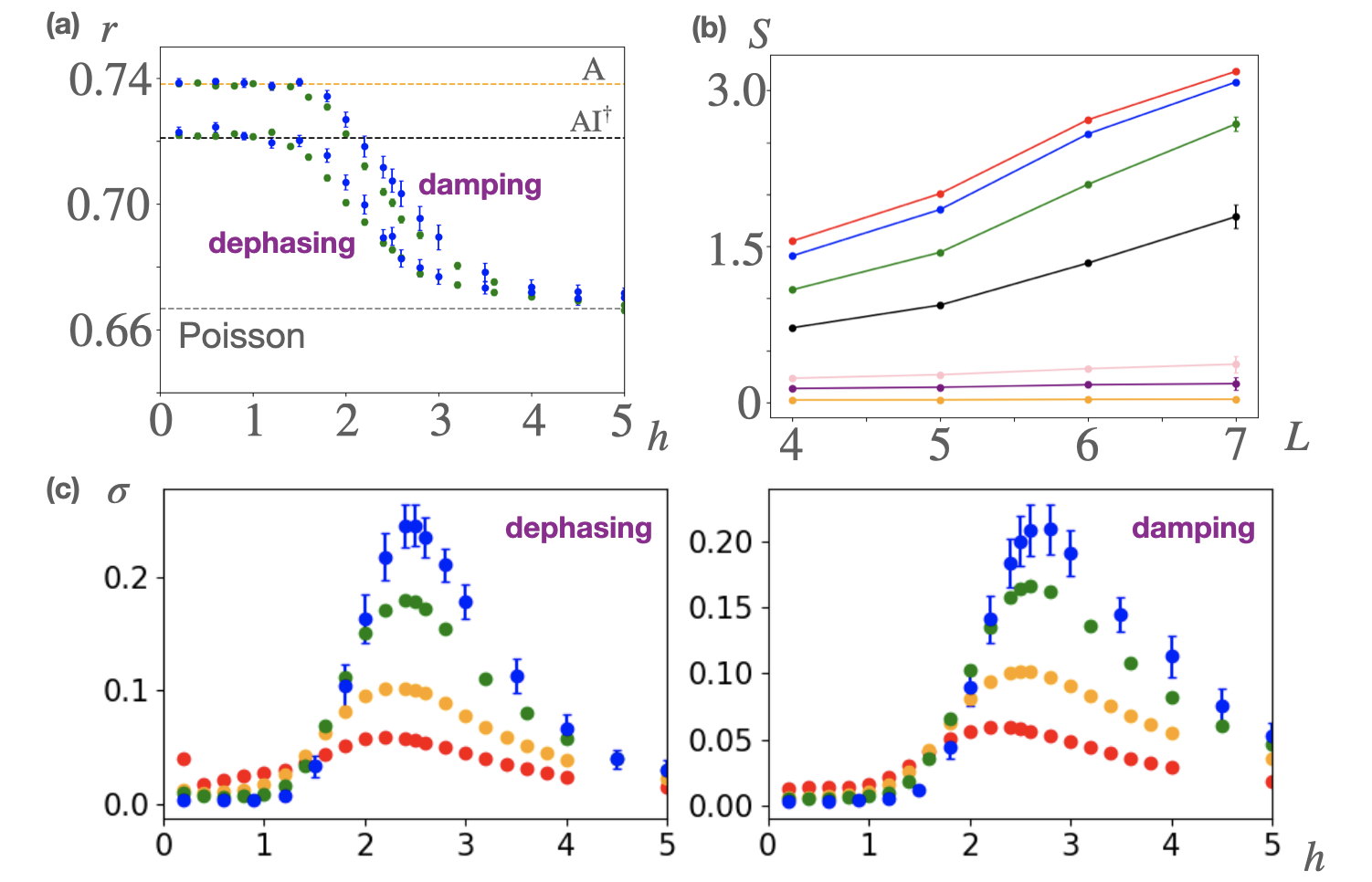}
\end{center}
\caption{
(a) Complex spacing ratio $r$ as a function of disorder strength $h$ with $L=6$ (green) and $L=7$ (blue) for  dephasing and damping types of dissipation.
For small $h$, $r$  becomes close to $r_\mr{AI^\dag}\simeq 0.721$ (dashed black line) and $r_\mr{A}\simeq 0.737$ (dashed orange line) for dephasing-type and  damping-type, respectively.
The value of $r$ decreases with increasing $h$, and becomes close to that for the Poisson distribution $r_\mr{Po}=2/3$ (dashed gray line).
(b) System-size dependence of OSEE $S_\alpha$ averaged over eigenstates in the middle of the spectrum for different disorder strength $h=0.6,1.2,1.8,2.4,4.0,5.0, 10.0$ (from red to orange).
While the OSEE increases with increasing $L$ for small $h$, the increase is suppressed for large $h$.
We  use the dephasing type of dissipation.
(c) Dependence on $h$ of the  variance $\sigma$  of the OSEE  for $4\leq L\leq 7$.
The peak develops as $L$ increases, from which the transition 
point reads as $h_c= 2.5\pm 0.1\:(2.7\pm 0.1)$ for dephasing (damping).
For (a)-(c), we use $J=1, g=-0.9,$ and $\gamma =0.5$ and the eigenvalues in the middle of the spectrum~\cite{Mid_foot}. 
The error bars indicate standard errors of the quantities evaluated from samples with different disorder realizations. 
The number of samples  is $10000~(L=4,5), 800~(L=6),$ and $50~(L=7)$.
}
\label{mbl}
\end{figure}

To strengthen evidence for the Lindbladian MBL transition, we calculate the spacing statistics of complex eigenvalues~\cite{Grobe88,Grobe89,Markum99,Akemann2019universal,hamazaki19nonhermitian,hamazaki2020universality,jaiswal2019universality,PhysRevLett.124.100604,sa2020complex,Sa20S,Tzortzakakis20,Yi20,mudute2020non,Luo21,Sa21I,rubio2021integrability,PhysRevA.105.L050201,PhysRevX.12.021040}, particularly the complex spacing ratio $r=\mbb{E}_\alpha\lrl{\lrv{\frac{\lambda_\alpha-\lambda_\alpha^\mr{N}}{\lambda_\alpha-\lambda_\alpha^\mr{NN}}}}$~\cite{sa2020complex}
as a function of $h$ in Fig.~\ref{mbl}(a), where
 $\lambda_\alpha^\mr{N}$ and $\lambda_\alpha^\mr{NN}$ are the nearest and the next-nearest neighbor eigenvalues of $\lambda_\alpha$ on the complex plane.
For small $h$, $r$ is close to $r_\mr{AI^\dag}\simeq 0.721$ and $r_\mr{A}\simeq 0.737$ for dephasing-type and damping-type, respectively.
Here, $r_\mr{AI^\dag}$ and $r_\mr{A}$ denote the complex spacing ratio for the non-Hermitian random matrices belonging to classes $\mr{AI}^\dag$ and $\mr{A}$.
On the other hand, $r$ decreases with increasing $h$ and eventually approaches the value for the Poisson distribution,  $r_\mr{Po}=2/3$~\cite{strong_foot}.
Note that Ref.~\cite{yusipov2021quantum} reports a similar eigenvalue transition for different disordered models, but with no mention to  eigenstates and the change in the dynamics associated with localization.

We next consider the operator-space entanglement  entropy (OSEE) $S_\alpha$~\cite{Prosen07,Pizorn09} of the (right) eigenstate $\hat{R}_\alpha$ for bipartition at the middle of the system $l=\lfloor L/2\rfloor$~\cite{OSEE_foot}.
Figure~\ref{mbl}(b) shows the system-size dependence of $S_\alpha$ averaged over eigenstates in the middle of the spectrum for different disorder strengths.
While the OSEE increases with increasing $L$ for small $h$, its shows a much slower increase for large $h$.
This is similar to the entanglement transition of the MBL in a closed system.

To find the transition point, we define the variance $\sigma$ of the OSEE with respect to the eigenstates and show its disorder-strength dependence for various $L$ (Fig.~\ref{dist}(c)).
The peak develops as we increase the system size, from which the transition 
point reads as $h_c= 2.5\pm 0.1\:(2.7\pm 0.1)$ for dephasing (damping).
The peak of $\sigma$ at the transition is  a dissipative counterpart of the peak of the variance of the entanglement entropy for the isolated MBL~\cite{Kjall14}.

\paragraph*{Discussion.}
We have discussed the transition for fixed $\gamma$ with $0<\gamma<g$.
In contrast, as detailed in Supplemental Material~\cite{Supp_foot}, by changing $\gamma$ (including $\gamma\geq g$), we numerically find that the critical value $h_c(\gamma)$ exhibits non-monotonic behavior. Namely, $h_c(\gamma)$ first increases and then decreases.
This implies that strong dissipation facilitates the Lindbladian MBL.
For sufficiently large $\gamma$, we find the breakdown of the non-Hermitian RMT statistics even for small $h$, as shown in Fig.~\ref{intro}.
This indicates that a localized regime appears even for the clean system; however, we leave it as a future problem to investigate this possibility for larger system sizes.

In the opposite limit $\gamma\ra 0$, it is nontrivial whether the Lindbladian MBL transition point $h_c(\gamma\ra 0)$ coincides with the Hermitian MBL transition point $h_{\mr{H},c}$.
We conjecture that these two transition points coincide under certain conditions (see Supplemental Material~\cite{Supp_foot}).

\paragraph*{Conclusion and outlook.}
We have demonstrated that localization of Lindbladian eigenstates can occur in the  open quantum many-body systems and stabilizes the decay rate of coherence without many-body decoherence despite interactions, if disorder is sufficiently strong.
The weakly disordered phase is characterized by the spectral statistics reflecting the universality of non-Hermitian RMT, and the strongly disordered phase is characterized by the Lindbladian MBL.

Our study raises interesting questions.
The nature of the transitions between the RMT and the Lindbladian MBL phases using larger system sizes needs to be investigated,  which may uncover, e.g., a new type of criticality defined by the spectral statistics of super-operators in dissipative systems.
It is also interesting to ask whether the delocalization-localization transitions of the spectrum found in this paper occur for models with different types of Hamiltonians (e.g., particle-number-conserving models) and dissipation (e.g., stochastic hopping~\cite{Diehl08,Yusipov17,Vakulchyk18,Haga21}).
Furthermore, it is intriguing to clarify the relation between our phases defined by the middle of the spectrum (relevant for transient-time dynamics) and the other types of MBL phenomenology under dissipation, such as the long-time thermalization~\cite{Fischer16,Levy16,Medvedyeva16I,Luschen16,Nieuwenburg17,morningstar2021avalanches,sels2021markovian} and the non-Hermitian MBL~\cite{hamazaki19nonhermitian}.

\begin{acknowledgements}
The numerical calculations were carried out with the help of QUSPIN~\cite{SciPostPhys.2.1.003,weinberg2019quspin}.
M.N. is supported by JSPS KAKENHI Grant No. JP20K14383.
T.H. is supported by JSPS KAKENHI Grant No. JP19J00525.
M.U. is supported by JSPS KAKENHI Grant No. JP22H01152.
\end{acknowledgements}

\bibliographystyle{apsrev4-1}
\bibliography{derh2.bib}

\begin{thebibliography}{120}
\bibitem [{pre()}]{prev_foot}%
  \BibitemOpen
  \href@noop {} {}\bibinfo {note}
  {References~\cite{morningstar2021avalanches,sels2021markovian} analyzed the
  Liouvillian gap of the Lindbladian super-operator, which is different from
  the middle of the spectrum of the Lindbladian considered in our work as an
  analogue with the Hermitian case.}\BibitemShut {Stop}%
\bibitem [{pol()}]{poly_foot}%
  \BibitemOpen
  \href@noop {} {}\bibinfo {note} {In addition to the exponential decay whose
  rate is proportional to $\gamma$, the coherence may have a polynomial decay
  factor due to interactions as in Hermitian
  cases~\cite{PhysRevB.96.014202,Abanin19}. However, this factor is negligible
  compared with the exponential decay.}\BibitemShut {Stop}%
\bibitem [{nor()}]{normalize_foot}%
  \BibitemOpen
  \href@noop {} {}\bibinfo {note} {We normalize the eigenstates such that
  $\Tr[\hat{R}_\alpha^\dag\hat{R}_\alpha]=\Tr[\hat{L}_\alpha^\dag\hat{L}_\alpha]=1$
  is satisfied except for the stationary states, for which
  $\Tr[\hat{R}_\mr{ss}]=1$ and $\hat{L}_\mr{ss}=\hat{\mbb{I}}$.}\BibitemShut
  {Stop}%
\bibitem [{rea()}]{realG_foot}%
  \BibitemOpen
  \href@noop {} {}\bibinfo {note} {While the analytical formula for ensembles
  different from a complex Ginibre ensemble (e.g., a real Ginibre ensemble and
  an ensemble with transposition symmetry~\cite{hamazaki2020universality}) is
  not known, we have numerically confirmed this distribution. If
  $\lambda_\alpha$ is real, then $r_\alpha^G$ and $r^{G'}_{\alpha}$ can be
  taken to be real, and their distributions may be different from the case of
  complex eigenvalues. We here do not consider real eigenvalues because their
  fractions vanish in the limit of large system size.}\BibitemShut {Stop}%
\bibitem [{pea()}]{peak_foot}%
  \BibitemOpen
  \href@noop {} {}\bibinfo {note} {The peaks at $x=0$ for
  $\Tr[\hat{L}_\alpha^\dag\hat{\rho}]$ and $\Tr[\hat{\mc{O}}\hat{R}_\alpha]$
  with large $h$ indicate that the eigenstates of the Lindbladian and those of
  $\hat{\rho}$ or $\hat{\mc{O}}$ can be orthogonal in the operator space for
  the localized regime.}\BibitemShut {Stop}%
\bibitem [{Mid()}]{Mid_foot}%
  \BibitemOpen
  \href@noop {} {}\bibinfo {note} {Specifically, let us consider the real part
  $\lambda_\alpha^r$ and the absolute value $|\lambda_\alpha^i|$ of the
  imaginary part of $\lambda_\alpha$, and sort each of them independently as
  $a_1\leq \cdots \leq a_{D^2}=0$ and $0\leq b_1\leq \cdots \leq b_{D^2}$. We
  use only those modes whose eigenvalues satisfy $a_{3D^2/8}\leq
  \lambda_\alpha^r\leq a_{5D^2/8}$ and $b_{3D^2/8}\leq |\lambda_\alpha^i|\leq
  b_{5D^2/8}$}\BibitemShut {NoStop}%
\bibitem [{OSE()}]{OSEE_foot}%
  \BibitemOpen
  \href@noop {} {}\bibinfo {note} {To define the OSEE, we consider a vector
  representation of $\hat{R}_\alpha$, i.e.,
  $|R_\alpha)=\sum_{ij}(\hat{R}_\alpha)_{ij}\ket{i}\otimes\ket{j}$, where
  $({R_\alpha|R_\alpha})=1$ because of the normalization condition. The entire
  system with the enlarged Hilbert space $\tilde{\mc{H}}=\mc{H}\otimes\mc{H}'$
  spanned by $\lrm{\ket{i}\otimes\ket{j}|\ket{i}\in\mc{H},\ket{j}\in\mc{H'}}$
  can be separated into two parts as
  $\tilde{\mc{H}}=\tilde{\mc{H}}_{AA}\otimes\tilde{\mc{H}}_{\av{AA}}$, where
  $\tilde{\mc{H}}_{AA}=\mc{H}_A\otimes \mc{H}_A'$ and
  $\tilde{\mc{H}}_{\av{AA}}=\mc{H}_{\av{A}}\otimes \mc{H}_{\av{A}}'$ with
  $\av{A}$ being the complement of the region $A$. Using the reduced density
  matrix
  $\hat{\mathtt{R}}_r=\Tr_{\tilde{\mc{H}}_{\av{AA}}}[|{R_\alpha})({R_\alpha}|]$,
  we define the OSEE as $
  S_\alpha=-\Tr[\hat{\mathtt{R}}_r\ln\hat{\mathtt{R}}_r] $}\BibitemShut
  {NoStop}%
\end{thebibliography}

\end{document}


\title{
Supplemental Material for ``Lindbladian Many-Body Localization"}
\date{\today}
\author{Ryusuke Hamazaki}
\affiliation{
Nonequilibrium Quantum Statistical Mechanics RIKEN Hakubi Research Team, RIKEN Cluster for Pioneering Research (CPR), RIKEN iTHEMS, Wako, Saitama 351-0198, Japan
}
\author{Masaya Nakagawa}
\affiliation{Department of Physics, University of Tokyo, 7-3-1 Hongo, Bunkyo-ku, Tokyo 113-0033, Japan}
\author{Taiki Haga}
\affiliation{Department of Physics and Electronics, Osaka Metropolitan University, Osaka, 599-8531, Japan}
\author{Masahito Ueda}
\affiliation{
Department of Physics, University of Tokyo, 7-3-1 Hongo, Bunkyo-ku, Tokyo 113-0033, Japan
}
\affiliation{
RIKEN Center for Emergent Matter Science (CEMS), Wako, Saitama 351-0198, Japan}
\affiliation{
Institute for Physics of Intelligence, University of Tokyo, 7-3-1 Hongo, Bunkyo-ku, Tokyo 113-0033, Japan}

\maketitle

\tableofcontents

\setcounter{equation}{0}
\setcounter{figure}{0}
\setcounter{section}{0}

\setcounter{table}{0}
\renewcommand{\theequation}{S-\arabic{equation}}
\renewcommand{\thefigure}{S-\arabic{figure}}
\renewcommand{\thetable}{S-\arabic{table}}

\section{Relaxation dynamics}
\subsection{Transverse and longitudinal relaxations}

\begin{figure}
\begin{center}
\includegraphics[width=\linewidth]{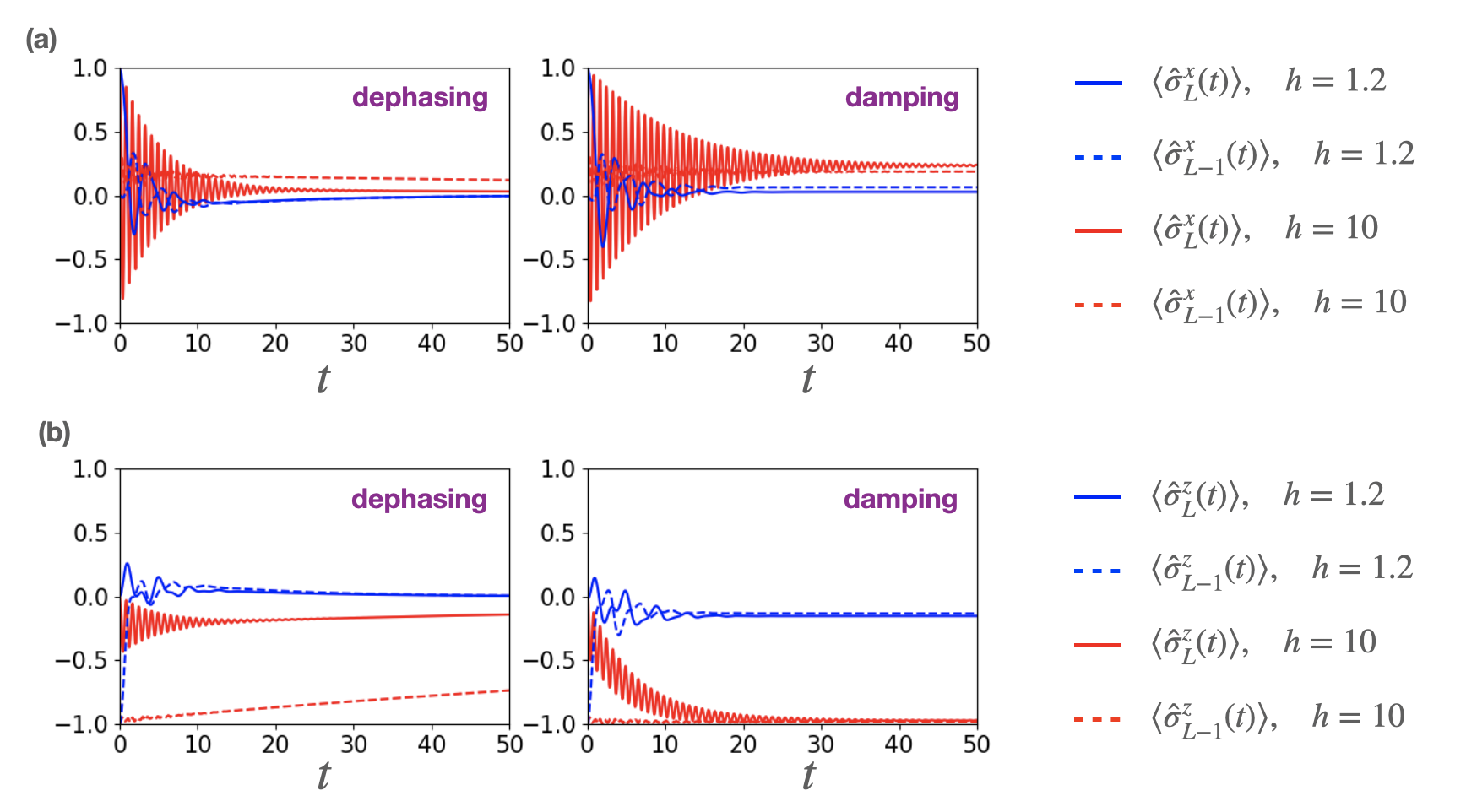}
\end{center}
\caption{
Transverse and longitudinal dynamics from an initial state given in Eq.~\eqref{initial} for the disordered Ising model with (left) dephasing and (right) damping types of dissipation.
(a) Transverse relaxation of spins. We consider the time evolutions of $\braket{\hat{\sigma}_{L}^x(t)}$ (solid) and $\braket{\hat{\sigma}_{L-1}^x(t)}$ (dashed) for $h=1.2$ (blue) and $h=10$ (red).
For $\braket{\hat{\sigma}_{L}^x(t)}$, we find rapid many-body decoherence for weak disorder and slower oscillatory decay (with time scale $\propto \gamma^{-1}$) for strong disorder.
On the other hand, $\braket{\hat{\sigma}_{L-1}^x(t)}$ does not show significant oscillations for all cases since the initial coherence at site $L-1$ is absent.
(b) Longitudinal relaxation of spins. We consider time evolutions of $\braket{\hat{\sigma}_{L}^z(t)}$ (solid) and $\braket{\hat{\sigma}_{L-1}^z(t)}$ (dashed) for $h=1.2$ (blue) and $h=10$ (red).
For dephasing, $\braket{\hat{\sigma}_{L-1}^z(t)}$ exhibits a slow decay.
Similarly, $\braket{\hat{\sigma}_{L}^z(t)}$ decays slowly, although small oscillation appears.
For the damping case, $\braket{\hat{\sigma}_{L-1}^z(t)}$ is almost constant around $-1$, and $\braket{\hat{\sigma}_{L}^z(t)}$ exhibits a decay to $-1$ with a time scale $\sim \gamma^{-1}$.
We use a single disorder realization and take $L=6, J=1, g=-0.9$, and $\gamma=0.1$.
}
\label{relaxation}
\end{figure}

Here, we numerically demonstrate the difference between longitudinal  and transverse (coherence) relaxation dynamics.
We consider the time evolution from the initial state with $q=1$ in the main text, i.e., 
\aln{\label{initial}
\ket{\psi_{q=1}}=(\ket{\downarrow}_1\cdots\ket{\downarrow}_{L-1})\lrs{\frac{\ket{\uparrow}_{L}+\ket{\downarrow}_{L}}{\sqrt{2}}}.
}

Figure~\ref{relaxation}(a) shows the dynamics of $\braket{\hat{\sigma}_{L}^x(t)}$ and $\braket{\hat{\sigma}_{L-1}^x(t)}$, which describe a transverse relaxation of spins.
For $\braket{\hat{\sigma}_{L}^x(t)}$, we find rapid many-body decoherence for weak disorder and slower oscillatory decay (with time scale $\propto \gamma^{-1}$) for strong disorder.
As discussed in the main text, the latter is attributed to the localization of off-diagonal degrees of freedom in the Lindbladian spectrum.
Note that $\braket{\hat{\sigma}_{L-1}^x(t)}$ does not show significant oscillations for all cases since the coherence at site  $L-1$ is absent from the beginning of the dynamics.

Figure~\ref{relaxation}(b) shows the dynamics of $\braket{\hat{\sigma}_{L}^z(t)}$ and $\braket{\hat{\sigma}_{L-1}^z(t)}$, which describe the longitudinal relaxation of spins.
For small $h$, we find a rapid relaxation to a stationary value for both dephasing and damping cases.
On the other hand, for strong disorder, the dynamics differ for these two different types of dissipation.
For dephasing, $\braket{\hat{\sigma}_{L-1}^z(t)}$ exhibits slow decay, 
which is consistent with previous studies~\cite{Fischer16,Levy16,Medvedyeva16I}.
This is attributed to the delocalization of diagonal degrees of freedom in the Lindbladian spectrum (especially around the zero eigenvalue). 
Similarly, $\braket{\hat{\sigma}_{L}^z(t)}$ decays slowly, although small oscillations appear due to the small overlap between $\hat{\sigma}_{L}^z$ and $\hat{\tau}_{L}^x$ (see Appendix IIIA for the definition of $\hat{\tau}_i^x$).
For the damping case, the stationary state is approximately described by a state in which spins are all down.
Then, $\braket{\hat{\sigma}_{L-1}^z(t)}$ is almost constant around $-1$, and $\braket{\hat{\sigma}_{L}^z(t)}$ exhibits a decay to $-1$ with a time scale $\sim (2\gamma)^{-1}$.
Note that this time scale for the longitudinal relaxation $\sim (2\gamma)^{-1}$ in the strong-disorder regime is twice as long as the time scale for the transverse relaxation $\sim (\gamma)^{-1}$ under damping-type dissipation.
While the factor two is known in the two-level dynamics described by a quantum-optical Master equation with damping~\cite{breuer2002theory}, it is surprising that the same factor appears even for many-body systems showing the Lindbladian MBL.



\subsection{Rigid decay rate for the dynamics of coherence}
\subsubsection{Rigid decay rate deep in the Lindbladian MBL}
Here, we numerically demonstrate that the decay rate for the dynamics of coherence approaches $\sim 2\gamma$ (dephasing) or 
$\sim \gamma$ (damping) for sufficiently strong disorder.
For this purpose, let us consider the dynamics of $\braket{\hat{\sigma}_L^x(t)}$ discussed in the previous subsection and fit it with the following function:
\aln{\label{fitting}
f(t)=e^{-at}(\cos bt+(1-c)\sin bt)
}
with fitting parameters $a,b$, and $c$.
The fitting function is motivated as follows:
(i) $\braket{\hat{\sigma}_L^x(0)}=1$, and (ii) an exponential decay of $\hat{\tau}_L^x$ is accompanied by an oscillation at an almost fixed frequency (corresponding to $2\tilde{h}_l$ in Eqs.~\eqref{eigdeph} or \eqref{eigdam}) in the deep Lindbladian MBL phase.
Note that the fitting may not work well for the delocalized phase.
Even in the Lindbladian MBL phase, the fitting may not work (e.g., an oscillation with multiple frequencies appears) due to the difference between $\hat{\tau}_l$ and $\hat{\sigma}_l$. 
Nevertheless, given that the fitting works well in the deep Lindbladian MBL phase with sufficiently strong disorder, its decay rate $a$ will be close to $\sim 2\gamma$ (dephasing) or 
$\sim \gamma$ (damping).

\begin{figure}
\begin{center}
\includegraphics[width=\linewidth]{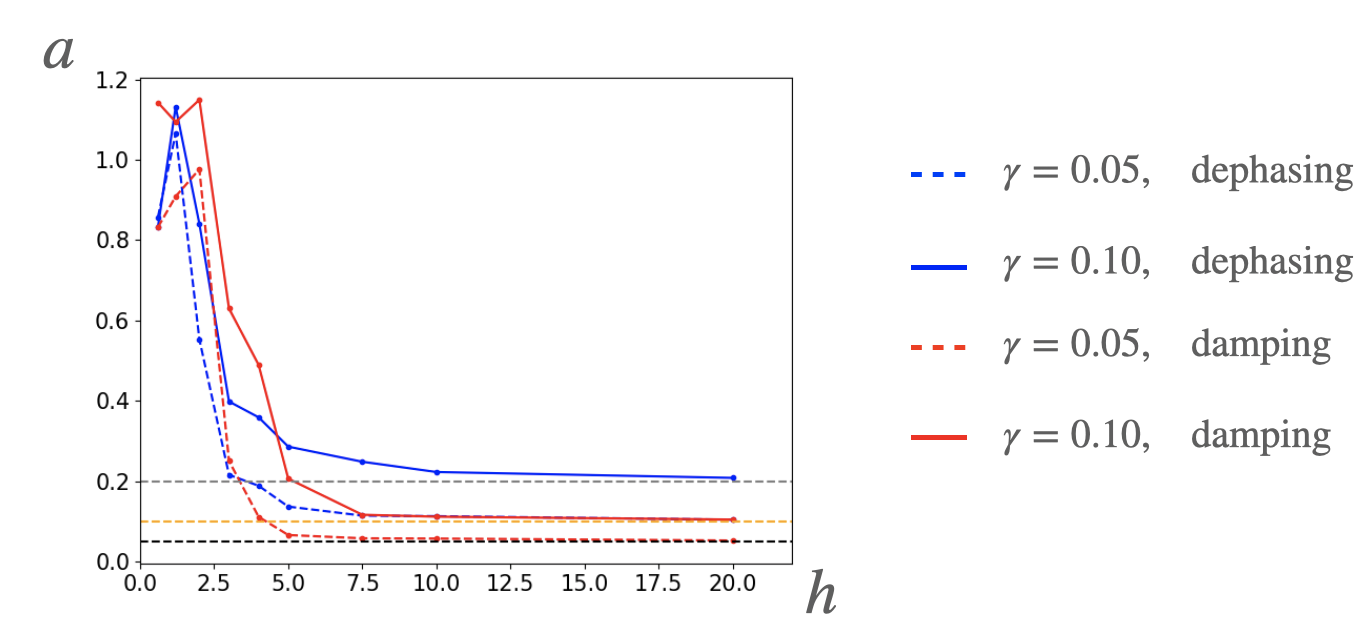}
\end{center}
\caption{
Fitting parameter $a$ averaged over the samples with $R\geq 0.9$, where we use $\gamma=0.05$ (dashed) and $\gamma=1.0$ (solid).
It decreases for $h\geq 2.0$ and approaches $\sim 2\gamma$ for dephasing (blue) or 
$\sim \gamma$ for damping (red) as $h$ is increased.
}
\label{fittingfig}
\end{figure}

Table~\ref{tb:fugafuga} shows the number of samples whose coefficient of determination $R$ for the fitting satisfies $R\geq 0.9$ among 100 samples.
For sufficiently large $h$, the fitting with Eq.~\eqref{fitting} becomes better for many samples, as expected.
Note that the fit seems to work well for small $h$ as well, but this is not due to localization, as the fitting parameter $a$ is far from $\gamma$.
Indeed, as shown in Fig.~\ref{fittingfig}, the fitting parameter $a$ averaged over the samples with $R\geq 0.9$ decreases for $h\geq 2.0$ and approaches $\sim 2\gamma$ (dephasing) or 
$\sim \gamma$ (damping) as $h$ is increased.

\begin{table}[htbp]
  \centering
  \begin{tabular}{c|cccccccccc}
     &\:\: $h=0.6$ \:\:& \:\:1.2 \:\:& \:\:2.0 \:\:& \:\:3.0\:\: & \:\:4.0\:\: &\:\: 5.0 \:\:& \:\:7.5\:\: &\:\:10.0 \:\:&\:\:20.0\:\:\\ \hline\hline
    dephasing ($\gamma=0.05$) & 4 & 16 & 11 & 12 & 29 & 43 & 55 &74 &80\\
    dephasing ($\gamma=0.1$) & 8 & 23 & 16 & 16 & 28 & 42 & 58 &69 &79\\ \hline
    damping ($\gamma=0.05$) & 18 & 31 & 12 & 8 & 13 & 27 & 46 &65 &77\\
    damping ($\gamma=0.1$) & 59 & 45 & 15 & 9 & 16 & 26 & 45 &62 &76\\ \hline
  \end{tabular}
  \caption{Number of samples whose coefficient of determination $R$ for the fitting satisfies $R\geq 0.9$ among 100 samples. For sufficiently large $h$, the fitting in Eq.~\eqref{fitting} becomes better for many samples. The fitting is made with the use of the SciPy \textsf{curve\_fit} function written in Python, and samples that fail to converge within the number of calls \textsf{maxfev=800} are neglected.}
  \label{tb:fugafuga}
\end{table}

\subsubsection{Comparison with the decay rate for a single-spin system}
Here, we compare the result in the previous subsection with the decay rate for a single-spin system.
For a single-spin system, the Lindblad equation reads
\aln{
\fracd{\hat{\rho}}{t}=-i[g\hat{\sigma}_1^x+h_1\hat{\sigma}_1^z,\hat{\rho}]+\frac{1}{2}\lrl{2\hat{\Gamma}_1\hat{\rho}\hat{\Gamma}_1^\dag-\lrm{\hat{\Gamma}_1^\dag\hat{\Gamma}_1,\hat{\rho}}}
}
with $\Gamma_1=\sqrt{\gamma}\hat{\sigma}_1^z$ for dephasing and
$\Gamma_1=\sqrt{\gamma/2}\hat{\sigma}_1^-$ for damping.

For simplicity, let us assume that $\gamma$ is small enough.
In that case,  the four eigenvalues of the Liouvillian super-operator are given by 
\aln{
0,\quad -\frac{2g^2}{g^2+h_1^2}\gamma +\mr{O}(\gamma^2),\quad -\frac{g^2+2h_1^2}{g^2+h_1^2}\gamma \pm 2\sqrt{g^2+h_1^2}i+\mr{O}(\gamma^2)
}
for dephasing,
and
\aln{
0,\quad -\frac{g^2+2h_1^2}{g^2+h_1^2}\gamma +\mr{O}(\gamma^2),\quad -\frac{3g^2+2h_1^2}{2(g^2+h_1^2)}\gamma \pm 2\sqrt{g^2+h_1^2}i+\mr{O}(\gamma^2)
}
for damping.

For the strong disorder $h_1\gg g$, the eigenvalues are approximated by
$\sim 0,0,2\gamma\pm 2h_1i$ for dephasing and  $\sim 0,-2\gamma,-\gamma\pm 2h_1 i$ for damping.
Thus, the longest lived modes have decay rates $2\gamma$ and $\gamma$ for dephasing and damping, respectively, as argued in the previous subsection and the main text.
We also note that the real parts for all the eigenvalues are larger than $-2\gamma$ for any $h_1$ and $g$.
This fact suggests that the large value of $a$ for small $h$ in the fitting to $f(t)$ (see Eq.~\eqref{fitting} and Fig.~\ref{fittingfig}) is not 
attributed to the decay rate for a single-spin system.

\subsubsection{Stability against additional interactions}
\begin{figure}
\begin{center}
\includegraphics[width=\linewidth]{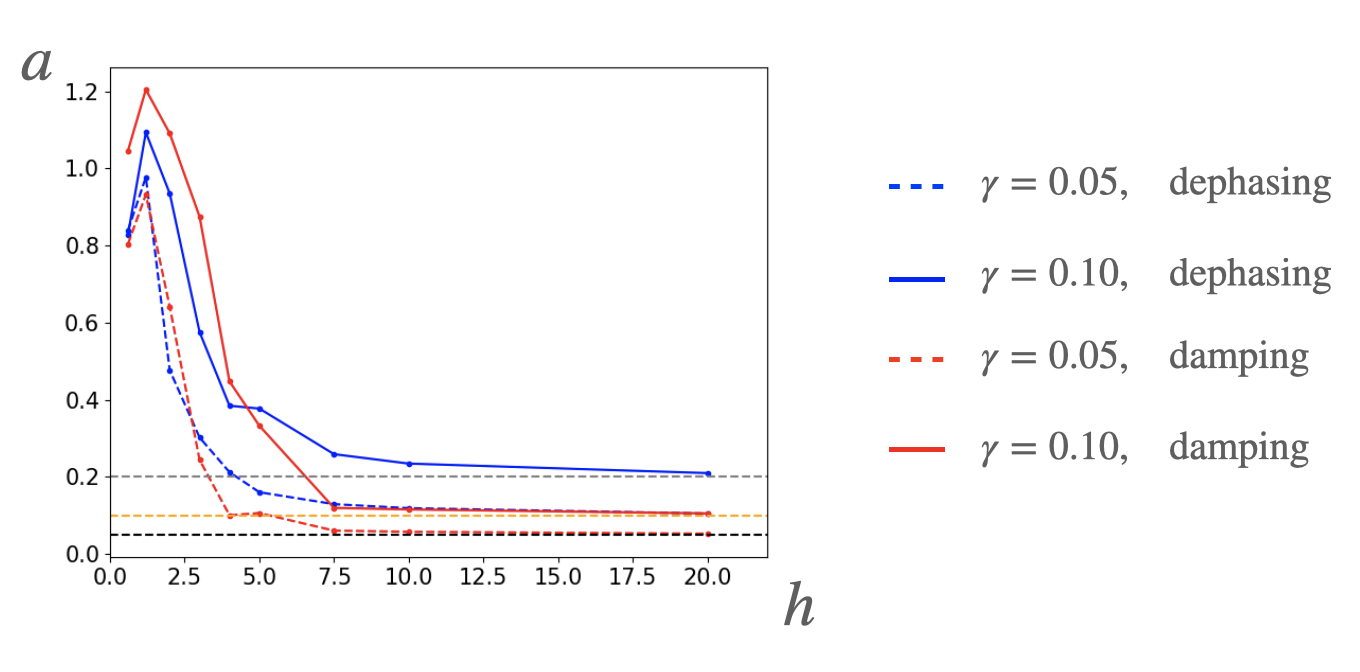}
\end{center}
\caption{
Fitting parameter $a$ averaged over the samples with $R\geq 0.9$ for the case where the Hamiltonian is perturbed by the XY interaction ($J_{XY}=0.2$).
Here, we use $\gamma=0.05$ (dashed) and $\gamma=1.0$ (solid).
The parameter decreases for $h\geq 2.0$ and approaches $\sim 2\gamma$ for dephasing (blue) or 
$\sim \gamma$ for damping (red) as $h$ is increased.
}
\label{fittingfigxy}
\end{figure}

To strengthen our argument, we demonstrate that the decay rate for large $h$ is robust against additional interactions distinct from the Ising one.
Here, we consider the case where the Hamiltonian is perturbed by an  XY-type interaction, i.e., 
\aln{
\hat{H}'=\hat{H}+J_{XY}\sum_{l=1}^{L-1} (\hat{\sigma}_l^x\hat{\sigma}_{l+1}^x+\hat{\sigma}_l^y\hat{\sigma}_{l+1}^y).
}
We can perform the same fitting procedure based on Eq.~\eqref{fitting} for the dynamics of coherence under the Hamiltonian $\hat{H}'$.
First, Table~\ref{tb:fugafuga2} shows the number of samples whose coefficient of determination $R$ for the fitting satisfies $R\geq 0.9$ among 100 samples.
Second, in Fig.~\ref{fittingfigxy}, we show the fitting parameter $a$ averaged over the samples with $R\geq 0.9$.
These two results are similar to the case without the XY interaction.
In particular, $a$ decreases for $h\geq 2.0$ and approaches $\sim 2\gamma$ (dephasing) or 
$\sim \gamma$ (damping) as $h$ is increased, which indicates the stability of the decay rate against perturbation.

\begin{table}[htbp]
  \centering
  \begin{tabular}{c|cccccccccc}
     &\:\: $h=0.6$ \:\:& \:\:1.2 \:\:& \:\:2.0 \:\:& \:\:3.0\:\: & \:\:4.0\:\: &\:\: 5.0 \:\:& \:\:7.5\:\: &\:\:10.0 \:\:&\:\:20.0\:\:\\ \hline\hline
    dephasing ($\gamma=0.05$) & 9 & 11 &4& 12 &27& 38 &53& 68 & 78\\
    dephasing ($\gamma=0.1$) & 14 & 22 &14& 20 &26& 39 & 54&67 & 77\\ \hline
    damping ($\gamma=0.05$) & 18 & 34 &8& 5 &13& 23 &42& 57 & 75\\
    damping ($\gamma=0.1$) & 58 & 57 &21 & 14 &15& 23 &41& 57 &74\\ \hline
  \end{tabular}
  \caption{Number of samples whose coefficient of determination $R$ for the fitting satisfies $R\geq 0.9$ among 100 samples for the case where the Hamiltonian is perturbed by the XY interaction ($J_{XY}=0.2$). 
  The fitting is made with the use of the SciPy \textsf{curve\_fit} function written in Python, and samples that fail to converge within the number of calls \textsf{maxfev=800} are neglected.}
  \label{tb:fugafuga2}
\end{table}

\section{Structures of $|\Tr[\hat{O} \hat{R}_\alpha]|, |\Tr[\hat{L}_\alpha^\dag\hat{\rho}]|$, and $|\Tr[ \hat{R}_\alpha^\dag\hat{L}_\alpha]|^{-1}$}

\subsection{System-size dependences of $\mc{C},\mc{B}_{\hat{\rho}},$ and $\mc{A}_{\hat{O}}$}

\begin{figure}
\begin{center}
\includegraphics[width=\linewidth]{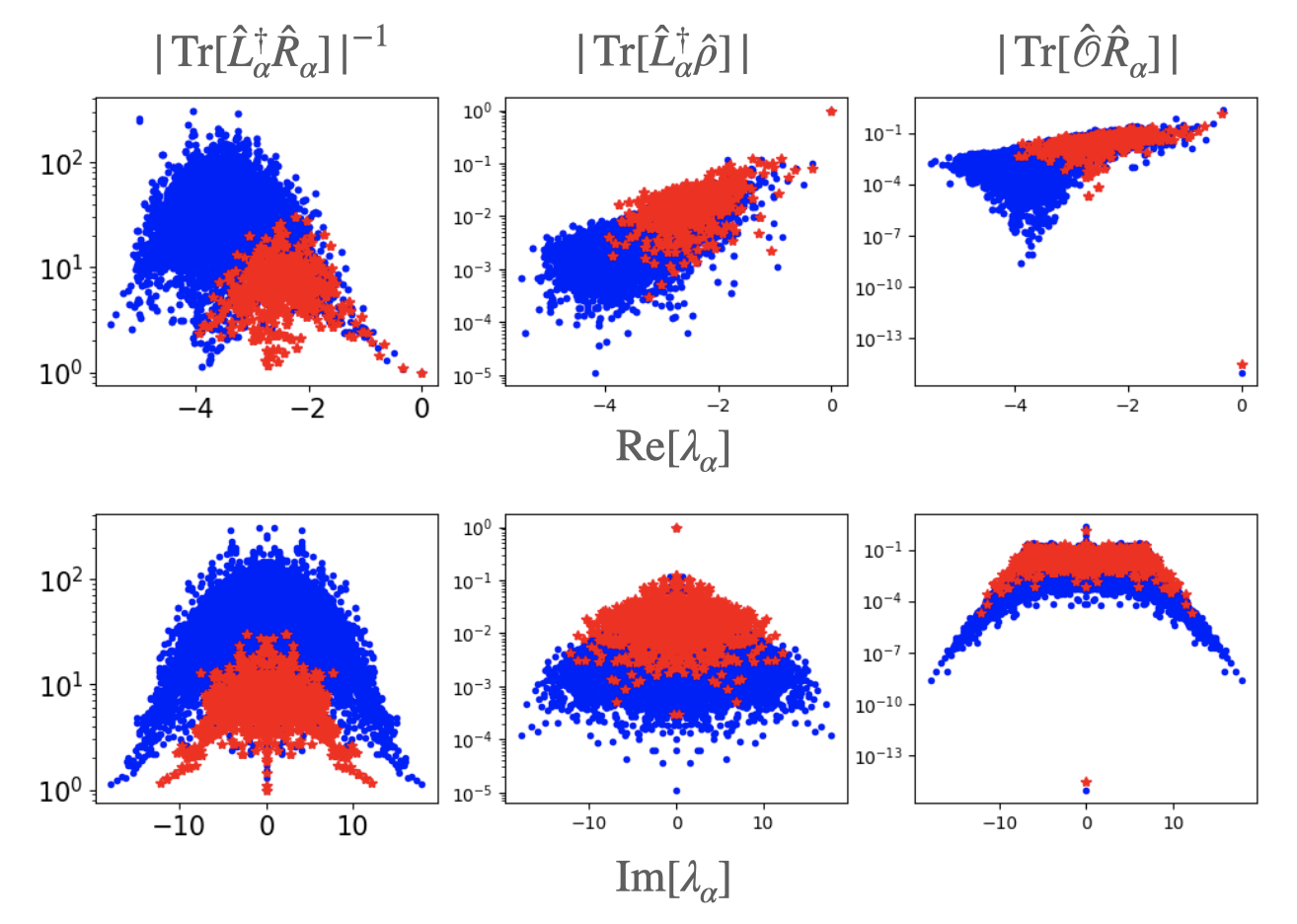}
\end{center}
\caption{Structures of $|\Tr[\hat{O} \hat{R}_\alpha]|, |\Tr[\hat{L}_\alpha^\dag\hat{\rho}]|$, and $|\Tr[ \hat{L}_\alpha^\dag\hat{R}_\alpha]|^{-1}$ for one sample, as a function of real and imaginary parts of $\lambda_\alpha$. 
While $|\Tr[\hat{O} \hat{R}_\alpha]|$ and $|\Tr[\hat{L}_\alpha^\dag\hat{\rho}]|$ tend to decrease with $L$, $|\Tr[ \hat{L}_\alpha^\dag\hat{R}_\alpha]|^{-1}$ increases with increasing $L$.
We consider $L=7$ (blue) and $L=5$ (red) for each plot.
We use the dephasing-type dissipation and set $q=1$, $J=1, g=-0.9, h=1.2$, and $\gamma =0.5$. 
}
\label{structures}
\end{figure}

\begin{figure}
\begin{center}
\includegraphics[width=\linewidth]{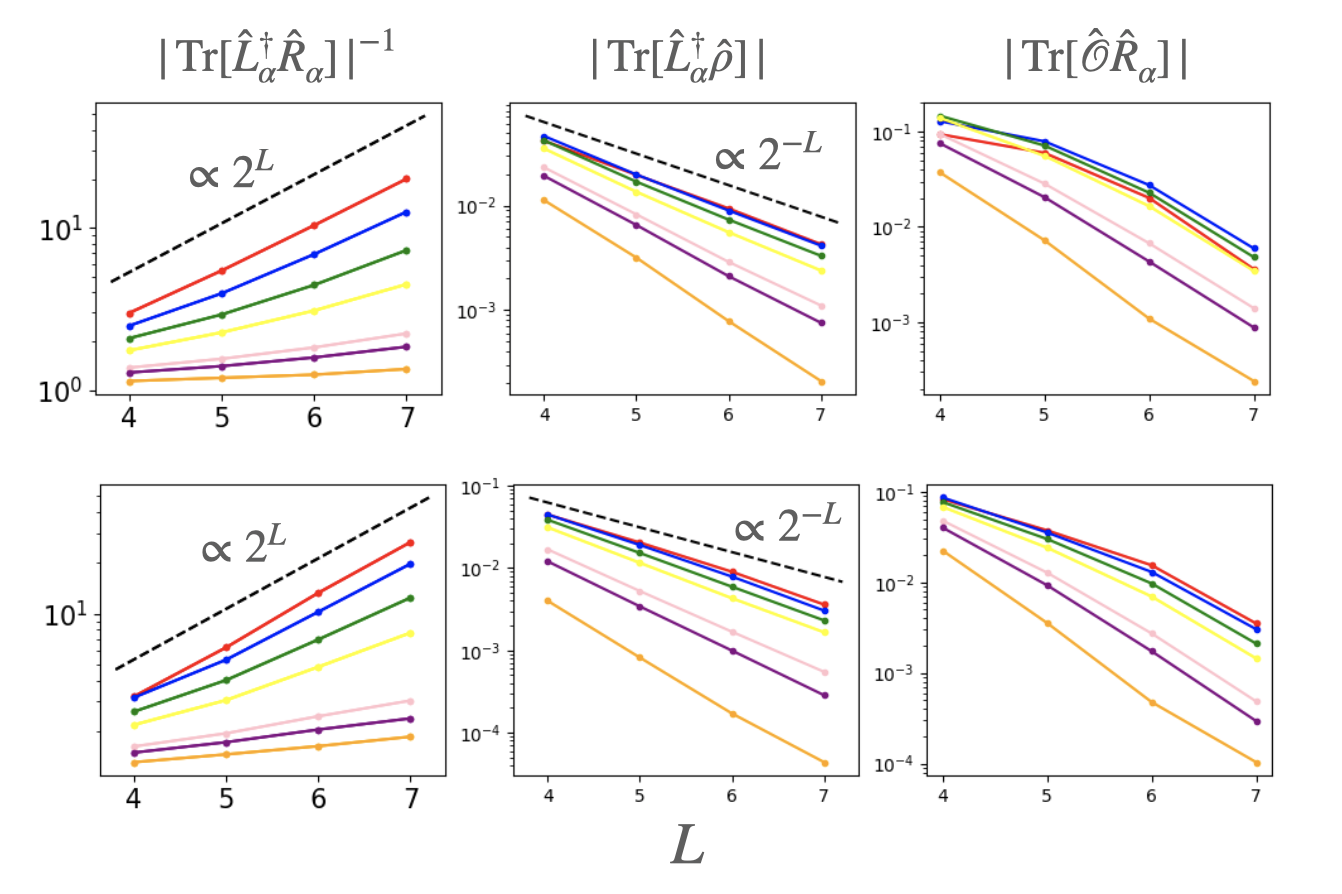}
\end{center}
\caption{Average values of $|\Tr[ \hat{L}_\alpha^\dag\hat{R}_\alpha]|^{-1}, |\Tr[ \hat{L}_\alpha^\dag\hat{\rho}]|,$ and $|\Tr[\hat{O}\hat{R}_\alpha]|$ for different values of $h$ ($=0.6,1.2,1.8,2.4,4.0,5.0, 10.0$ from red to orange) and for dephaing (top) and  damping (bottom) types of dissipation.
We see that  $|\Tr[ \hat{L}_\alpha^\dag\hat{R}_\alpha]|^{-1}\propto D=2^L$ and $\Tr[\hat{L}_\alpha^\dag\hat{\rho}]\propto D^{-1}$ for small $h$.
We use the Lindbladian with $L=7, J=1, g=-0.9$, and $\gamma =0.5$.
For the calculation of the average, we consider eigenvalues within the range $\mr{Re}[\lambda_\alpha]\in [0.6\lambda^r_\mr{min},0.4\lambda^r_\mr{min}]$ and $\mr{Im}[\lambda_\alpha]\in [0.6\lambda^i_\mr{min},0.4\lambda^i_\mr{min}]$, where $\lambda^r_\mr{min}=\min_\alpha\mr{Re}[\lambda_\alpha]$ and $\lambda^i_\mr{min}=\min_\alpha\mr{Im}[\lambda_\alpha]$. 
The number of samples used for the simulation is $10000 \:(L=4,5), 800\: (L=6),$ and $50\: (L=7)$.
}
\label{scaling1}
\end{figure}

We here discuss the system-size dependences of  $\mc{C},\mc{B}_{\hat{\rho}},$ and $\mc{A}_{\hat{O}}$ (see Eqs. (3)-(5) in the main text) in the middle of the spectrum.
Figure~\ref{structures} shows  $|\Tr[\hat{O} \hat{R}_\alpha]|, |\Tr[\hat{L}_\alpha^\dag\hat{\rho}]|$, and $|\Tr[ \hat{R}_\alpha^\dag\hat{L}_\alpha]|^{-1}$  for one sample for system sizes $L=5$ and $L=7$.
While $|\Tr[\hat{O} \hat{R}_\alpha]|$ and $|\Tr[\hat{L}_\alpha^\dag\hat{\rho}]|$ tend to decrease with $L$, $|\Tr[ \hat{L}_\alpha^\dag\hat{R}_\alpha]|^{-1}$ increases for increasing $L$.
Figure~\ref{scaling1} shows finite-size scaling of these factors averaged over many samples, where the disorder strength $h$ is varied.
We find that
 $|\Tr[ \hat{L}_\alpha^\dag\hat{R}_\alpha]|^{-1}$ is almost proportional to $D=2^L$ for small $h$,
where $D$ is the dimension of the Hilbert space.
In addition, we also find $|\Tr[\hat{L}_\alpha^\dag\hat{\rho}]|\propto D^{-1}$ for small $h$.
These scalings agree with the prediction of the non-Hermitian RMT (see Appendix IIB). 
On the other hand, for strong $h$, the scaling behavior differs from the RMT for both cases.
For $|\Tr[ \hat{O}\hat{R}_\alpha]|$, the scaling behavior is not straightforward because of the locality of the operator and that of the Lindblad generator (see Appendix IIC).

\subsection{Non-Hermitian random-matrix prediction of eigenstates}~\label{nonHermRMT}

The expressions in Eqs.~(3), (4), and (5) in the main text are motivated by the non-Hermitian random-matrix theory: Let $\ket{R_\alpha^G}/\ket{L_\alpha^G}$ be right/left eigenstates of an  $N\times N$ non-Hermitian real Ginibre random matrix.
Then we obtain
\aln{\label{RMT}
\braket{\mc{O}|R_\alpha^G},\:\braket{L_\alpha^G|\mc{O}}\sim \frac{\sqrt{\braket{\mc{O}|\mc{O}}}}{\sqrt{N}}r^G_{\alpha}
}
and
\aln{\label{northo}
\braket{L_\alpha^G|R_\alpha^G}^{-1}\sim \sqrt{N}r^{G'}_{\alpha},
}
where $\ket{\mc{O}}=(\bra{\mc{O}})^\dag$ is an arbitrary vector with dimension $N$, and $r^G_\alpha$ and $r^{G'}_\alpha$ are random variables with zero mean and unit variance.
The non-orthogonality of eigenstates in Eq.~$\eqref{northo}$ was studied previously in Refs.~\cite{chalker1998eigenvector,bourgade2019distribution,mehlig2000statistical,janik1999correlations,fyodorov2018statistics}.

Let us assume that $\lambda_\alpha$ has a nonzero imaginary part.
In this case, we find that the phases of $r_\alpha^G$ and $r^{G'}_{\alpha}$ are  randomly distributed over $[0,2\pi)$.
The squared absolute value of $r_\alpha^G$, $|r_\alpha^G|^2$, is essentially given by the Porter-Thomas distribution. 
From this we can obtain
\aln{
P_G(x):=P(|r_\alpha^G|=x)=\frac{\pi}{2}xe^{-\frac{\pi}{2}x^2}
}
after a normalization to set the average of $|r_\alpha^G|$ unity.
Moreover, $P(|r^{G'}_\alpha|=x)$ becomes a universal distribution for non-Hermitian random matrices, and it is expected to take the form of
\aln{
P_G'(x):=P(|r_\alpha^{G'}|=x)=\frac{32}{\pi^2 x^5}e^{-\frac{4}{\pi x^2}}.
}
Note that this function is originally obtained from the analytical results of  a complex Ginibre ensemble~\cite{bourgade2019distribution}.
While the analytical formula for a real Ginibre ensemble is not known, we have numerically confirmed this distribution.
If $\lambda_\alpha$ is real, then $r_\alpha^G$ and $r^{G'}_{\alpha}$ can be taken to be real, and their distributions may be different from the case of complex eigenvalues.

Now, we note that any matrix can have a vector representation through the vectorization mapping (see, e.g., Ref.~\cite{PhysRevB.99.214306})
\aln{\label{map}
\hat{A}=\sum_{ij}(\hat{A})_{ij}\ket{i}\bra{j} \rightarrow |A)=\sum_{ij}(\hat{A})_{ij}\ket{i}\otimes\ket{j}.
}
Using this representation, we have a form similar to the left-hand sides of Eqs.~\eqref{RMT} and \eqref{northo}:
\aln{
\Tr[\hat{O}\hat{R}_\alpha]=({O}|{R}_\alpha),
}
\aln{
\Tr[\hat{L}_\alpha^\dag\hat{\rho}]=(L_\alpha|\rho),
}
and
\aln{
\Tr[\hat{L}_\alpha^\dag\hat{R}_\alpha]=({L}_\alpha|{R}_\alpha),
}
where $(A|=\{|A)\}^\dag$.
Then, in analogy with the eigenstate thermalization hypothesis, which states that eigenstates of generic Hamiltonians obey the statistics of a  random matrix, 
we conjecture that Eqs.~(3), (4) and (5) in the main text hold for generic super-operators with $r_\alpha$ and $r_\alpha'$ obeying the distributions $P_G$ and $P_G'$, respectively.
Furthermore, owing to the $N$-scaling in Eq.~\eqref{northo}, we can also conjecture that $\mc{C}(\lambda_\alpha)\propto \sqrt{D^2}=D$ (note that $|R_\alpha)$ and the other vectorized states have dimension $D^2=2^{2L}$).
Numerical simulations indicate that the similar scaling also occurs for $\Tr[\hat{L}_\alpha^\dag\hat{\rho}]\sim D^{-1}$ (especially for pure states, where $(\rho|\rho)=1$) but not for $\Tr[\hat{O}\hat{R}_\alpha]$ owing to the locality of our model (see Appendix~IIC).

\subsection{Effect of locality on  $\Tr[\hat{O}\hat{R}_\alpha]$}\label{Local}

We here argue that $|\Tr[\hat{O}\hat{R}_\alpha]|$ is exponentially suppressed as a function of $|\lambda_\alpha|$ for a local Lindbladian and a local observable $\hat{O}$, unless many of $|R_\alpha)$ are almost parallel.
To explain the idea behind this argument, we consider 
\aln{
\sum_{|\lambda_\alpha|>\lambda}|\Tr[\hat{O}\hat{R}_\alpha]|^2
}
for some positive $\lambda$.
Using $\mc{L}[\hat{R}_\alpha]=\lambda_\alpha\hat{R}_\alpha$, we find
\aln{
|\Tr[\hat{O}\hat{R}_\alpha]|^2 =\frac{|\Tr[\hat{O}\mc{L}^k[\hat{R}_\alpha]]|^2}{|\lambda_\alpha|^{2k}}
=\frac{|\Tr[\mc{L}^{\dag k}[\hat{O}]\hat{R}_\alpha]|^2}{|\lambda_\alpha|^{2k}}
}
for an arbitrary positive integer $k$.
Here, by taking $\mc{L}^{\dag k}[\hat{O}]=:\hat{A}=\hat{A}^\dag$, we have 
\aln{
\sum_\alpha|\Tr[\mc{L}^{\dag k}[\hat{O}]\hat{R}_\alpha]|^2
&=\sum_\alpha\sum_{\mu\nu\xi\eta}A_{\mu\nu}^*R^\alpha_{\mu\nu}R_{\xi\eta}^{\alpha *}A_{\xi\eta}\nonumber\\
&=\sum_{\mu\nu\xi\eta}A_{\mu\nu}^*X_{\mu\nu;\xi\eta}A_{\xi\eta}\nonumber\\
&\leq \sum_{\mu\nu}|A_{\mu\nu}|^2 ||X||_\mr{OP}=\Tr[\hat{A}^2]||X||_\mr{OP}\leq D||\hat{A}||_\mr{\infty}^2||X||_\mr{OP},
}
where
\aln{
X_{\mu\nu;\xi\eta}={\sum_\alpha R^\alpha_{\mu\nu}R_{\xi\eta}^{\alpha *}}
}
and $||X||_\mr{OP}$ is defined as the operator norm for this $D^2\times D^2$ matrix ($||\hat{A}||_\mr{\infty}$ is the operator norm for the $D\times D$ matrix).
We therefore have
\aln{
\sum_{|\lambda_\alpha|>\lambda}|\Tr[\hat{O}\hat{R}_\alpha]|^2
&\leq\sum_{|\lambda_\alpha|>\lambda}\frac{|\Tr[\mc{L}^{\dag k}[\hat{O}]\hat{R}_\alpha]|^2}{|\lambda|^{2k}}\nonumber\\
&\leq\sum_{\alpha}\frac{|\Tr[\mc{L}^{\dag k}[\hat{O}]\hat{R}_\alpha]|^2}{|\lambda|^{2k}}\nonumber\\
&\leq D||X||_\mr{OP}\frac{||\mc{L}^{\dag k}[\hat{O}]||^2_\infty}{|\lambda|^{2k}}.
}

Next, if  $\mc{L}$ is a local super-operator and $\hat{O}$ is a local operator, 
we can show that $||\mc{L}^{\dag k}[\hat{O}]||_\infty$ can be bounded  from above, using a technique similar to the case of the  Hamiltonian formalism~\cite{mori2016rigorous}.
For simplicity, we assume that $\mc{L}$ and $\hat{O}$ are at most $s$-local. Then we have
\aln{
||\mc{L}^{\dag k}[\hat{O}]||_\infty\leq o_l(2\ell_lks)^k,
}
where $o_l$ and $\ell_l$ represent the norms of the local operators, which are  independent of the system size.
Then, by choosing $k=\lambda/2\ell_l se$, we have
\aln{\label{bound}
\sum_{|\lambda_\alpha|>\lambda}|\Tr[\hat{O}\hat{R}_\alpha]|^2
&\leq  Do_l^2 e^{-\frac{\lambda}{\ell_l se}}||X||_\mr{OP}.
}
Note that, for the Hermitian case, $\hat{R}_\alpha=\ket{E_a}\bra{E_b}\:(\alpha=(a,b))$ and thus $X_{\mu\nu;\xi\eta}=\delta_{\mu\xi}\delta_{\nu\eta}$, and hence  $||X||_\mr{OP}=1$.
However, owing to non-Hermiticity, $||X||_\mr{OP}$ can be larger; if we assume that all  the $|R_\alpha)$ are almost parallel, we have $||X||_\mr{OP}\sim D^2$.
In this case, the right-hand side of Eq.~\eqref{bound} becomes large, and the bound may be trivial.
On the other hand, we numerically find that $||X||_\mr{OP}$ does not increase faster than  $\propto {D}$ for our models.
Thus, assuming this scaling and noting that the number of elements of the order of $D^2$ contributes to the sum in the left-hand side, our rough estimation becomes
\aln{\label{expdecay}
|\Tr[\hat{O}\hat{R}_\alpha]|\lesssim o_l e^{-\frac{|\lambda_\alpha|}{2\ell_l se}}.
}
Due to the presence of the factor  $e^{-\frac{|\lambda_\alpha|}{2\ell_l se}}$ on the right-hand side, $|\Tr[\hat{O}\hat{R}_\alpha]|$ is suppressed in the middle of the spectrum for our models, where $\mr{Re}[\lambda_\alpha]$ is proportional to the volume of the system.
Thus, eigenmodes in the middle of the spectrum do not contribute to the dynamics of local observables.
Note that this is similar to the results in Refs.~\cite{PhysRevLett.124.100604,PhysRevResearch.3.023190}, which focus on the local random Lindbladian without Hamiltonian terms.

\subsection{Similar hypothesis for classical dynamics}\label{classical}
Here, we briefly discuss that eigenstates of classical stochastic systems (i.e., eigenstates of a Markovian generator) can be described by random matrices as in the Lindblad case.
For a general classical Markovian dynamics, we consider a $D\times D$ non-Hermitian matrix $M$ as a generator of the dynamics.
We then obtain $\vec{p}(t)=e^{Mt}\vec{p}(0)$, where $\vec{p}=(p_1,\cdots, p_D)^{\sf{T}}$ is the probability vector in the state space.
By the spectral decomposition
\aln{
M=\sum_a \lambda_a\frac{\vec{R}_a\vec{L}_a^\dag}{(\vec{R}_a,\vec{L}_a)},
}
where $(\vec{a},\vec{b})$ denotes the inner product and $\mr{Re}[\lambda_a]\leq 0$,
the time evolution for an observable $\vec{\mc{O}}$ can be written as 
\aln{
\braket{\mc{O}(t)} =(\vec{\mc{O}},\vec{p}(t))=\sum_a e^{\lambda_a t}\frac{(\vec{\mc{O}},\vec{R}_a)(\vec{L}_a,\vec{p})}{(\vec{R}_a,\vec{L}_a)}.
}
For a sufficiently complex dynamics, we conjecture that  assumptions similar to the Lindblad case hold true, namely
\aln{\label{noent}
(\vec{\mc{O}},\vec{R}_a)\sim \mc{A}_{\vec{\mc{O}}}(\lambda_a)r_\alpha,
}
\aln{\label{noent}
(\vec{\mc{O}},\vec{R}_a)\sim \mc{B}_{\vec{p}}(\lambda_a)r_\alpha
}
and 
\aln{
(\vec{L}_a,\vec{R}_a)^{-1}\sim \mc{C}(\lambda_a)r'_\alpha.
}
We note that $r_\alpha$ and $r_\alpha'$ obey non-Hermitian random matrix theory but $\mc{C}(\lambda_a)$ will be proportional to $\sqrt{D}$ rather than $D$. 
In addition, employing a derivation similar to that for Eq.~\eqref{expdecay} in the Lindblad case, we conjecture that for local $M$ and $\vec{\mathcal{O}}$,
\aln{
\mc{A}_{\vec{\mc{O}}}(\lambda_a)\propto e^{-\mr{O}(|\lambda_a|)}
}
for a locally interacting classical Markovian generator $M$.

\section{Phenomenology of the Lindbladian MBL phase}\label{embl}
\subsection{Structure of the Lindblad generator}
To analyze the effects of dissipation in the MBL phase, we consider the matrix representation of the Lindbladian as in Appendix~II.
In accordance with the vector representation of the matrix in Eq.~\eqref{map}, the Lindbladian super-operator is mapped to an operator
\aln{
\tilde{\mc{L}}=-i(\hat{H}\otimes\mbb{\hat{I}}-\hat{\mbb{I}}\otimes\hat{H}^T)+\sum_l\lrs{\hat{\Gamma}_l\otimes\hat{\Gamma}_l^*-\frac{1}{2}\hat{\Gamma}_l^\dag\hat{\Gamma}_l\otimes\hat{\mbb{I}}-
\frac{1}{2}\hat{\mbb{I}}\otimes\hat{\Gamma}_l^T\hat{\Gamma}_l^*}.
}
In the fully MBL phase, the Hamiltonian part can be diagonalized by a quasi-local unitary operator $U$ as~\cite{Huse14}
\aln{
\hat{H}=\sum_l \tilde{h}_l\hat{\tau}_l^z+\sum_{lm}J_{lm}\hat{\tau}_l^z\hat{\tau}_m^z+
\sum_{lmn}J_{lmn}\hat{\tau}_l^z\hat{\tau}_m^z\hat{\tau}_n^z+\cdots,
}
where the coupling constants $J_{lm},J_{lmn},\cdots$ decay exponentially as the distance between any two sites increases.
The local integral of motion $\hat{\tau}_l^z=U\hat{\sigma}_l^zU^\dag$ has a large overlap with $\hat{\sigma}_l^z$:
\aln{
\hat{\tau}_l^z=\sum_{\alpha=x,y,z}B_l^\alpha\hat{\sigma}_l^\alpha+\sum_{jk}\sum_{\alpha,\beta=x,y,z}D_{l,jk}^{\alpha\beta}\hat{\sigma}_j^\alpha\hat{\sigma}_k^\beta+\text{ (higher order terms)},
}
where $D_{l,jk}^{\alpha\beta}$ decays exponentially with respect to the distances $|l-j|$ and $|l-k|$.
For the case of the  transverse- and longitudinal-field Ising model, the first-order Schrieffer-Wolff transformation~\cite{Imbrie16D,Imbrie16O} suggests that $B_l^z=1-\mr{O}((g/h)^2)$, $B_l^{x,y}=\mr{O}(g/h)$ and $D_{l,jk}^{\alpha\beta}=\mr{O}((g/h)^2)$, if the rare resonant sites are neglected.
Conversely, we can expand $\hat{\sigma}_l^z=U^\dag\hat{\tau}_l^z U$ with the product of $\hat{\tau}_l^\alpha$, where $\hat{\tau}_l^x=U\hat{\sigma}_l^x U^\dag$ and $\hat{\tau}_l^y=U\hat{\sigma}_l^y U^\dag$, as
\aln{\label{LIOMsig}
\hat{\sigma}_l^z=\sum_{\alpha=x,y,z}Z_l^\alpha\hat{\tau}_l^\alpha+\sum_{jk}\sum_{\alpha,\beta=x,y,z}G_{l,jk}^{\alpha\beta}\hat{\tau}_j^\alpha\hat{\tau}_k^\beta+\text{(higher order terms)},
}
where $G_{l,jk}^{\alpha\beta}$ decays exponentially with respect to the distances $|l-j|$ and $|l-k|$, $Z_l^z=1-\mr{O}((g/h)^2)$, $Z_l^{x,y}=\mr{O}(g/h)$, and $G_{l,jk}^{\alpha\beta}=\mr{O}((g/h)^2)$.
A similar expression holds for $\hat{\sigma}_l^-$.

\subsection{Case of dephasing}
Let us first consider the case of dephasing.
We assume that the dissipation strength $\gamma$ is not so large.
We can decompose the matrix representation of the Lindbladian as
\aln{
\tilde{\mc{L}}=-i(\hat{H}\otimes\mbb{\hat{I}}-\hat{\mbb{I}}\otimes\hat{H}^T)+\sum_l({Z_l^z})^2\gamma(\hat{\tau}_l^z\otimes\hat{\tau}_l^z-\hat{\mbb{I}}\otimes\hat{\mbb{I}})+\tilde{\mc{L}}_P,
}
where $\tilde{\mc{L}}_P$ is a perturbation term of the order of  $g/h$, which arises from the second- or higher-order terms in Eq.~\eqref{LIOMsig}.
Without $\tilde{\mc{L}}_P$, the eigenstates are written as 
 \aln{
|{R}_\alpha)=|{L}_\alpha)=\ket{\tau_1\cdots\tau_L}\otimes \ket{\tau'_1\cdots\tau'_L}=|\phi^{\alpha_1}_1)|\phi^{\alpha_2}_2)\cdots |\phi^{\alpha_L}_L).
 }
Here, $|\phi^{\alpha_l}_l)=|\tau_l,\tau_l')=\ket{\tau_l}\otimes\ket{\tau_l'}$ is the vector representation of $\hat{\phi}_l^{\alpha_l}=\ket{\tau_l}\bra{\tau_l'}$, where $\ket{\tau_l}$ is the eigenstate of $\hat{\tau}_l^z$ with eigenvalue $\tau_l\:(=\pm 1)$.
Specifically, we take $|\phi^1)=|+1,-1)$, $|\phi^2)=|-1,+1)$, $|\phi^3)=|+1,+1)$, and 
$|\phi^4)=|-1,-1)$.
The corresponding eigenvalue of $|R_\alpha)$ is given by
\aln{
\lambda_\alpha=&-i\lrs{\sum_l \tilde{h}_l{\tau}_l+\sum_{lm}J_{lm}{\tau}_l{\tau}_m+\cdots}
+i\lrs{\sum_l \tilde{h}_l{\tau}_l'+\sum_{lm}J_{lm}{\tau}_l'{\tau}_m'+\cdots}\nonumber\\
&+\sum_l({Z_l^z})^2\gamma({\tau}_l{\tau}_l'-1),
}
which can also be written as
\aln{\label{eigdeph}
\lambda_\alpha=-i\sum_l2\tilde{h}_l(\delta_{\alpha_l,1}-\delta_{\alpha_l,2})-i(\text{small terms including } J_{lm}, \cdots)-\sum_l2{Z_l^z}^2\gamma(\delta_{\alpha_l,1}+\delta_{\alpha_l,2}).
}

Let us consider the effect of $\tilde{\mc{L}}_P$.
This perturbation includes the bit flip terms, i.e., terms that can flip $\ket{+1}\:(\ket{-1})$ (or its product) to $\ket{-1}\:(\ket{+1})$ (or its product).
To simplify the notation, we write $|\phi^{\alpha_1}_1)|\phi^{\alpha_2}_2)\cdots |\phi^{\alpha_L}_L)$ as $|\alpha_1\alpha_2\cdots\alpha_L)$.
First,  states such as $|d_1\cdots d_L)\:(d_l=3,4)$ [$d$ stands for diagonal degree of freedom] are unstable under perturbation and mixed with the other states in the form $|d'_1\cdots d'_L)\:(d_l'=3,4)$, since the energy difference between these states is zero.
Note that these states correspond to the long-time  longitudinal relaxation of a local integral of motion, which can be treated by an effective classical rate  equation~\cite{Fischer16,Medvedyeva16I}. 
Next, consider a state that contains a large cluster of $d_l$ with size $c\gg 1$, e.g., $[d_i\cdots d_k\cdots d_{i+c-1}]\:(d_l=3,4)$ with $|i-k|, |i+c-1-k|\gg 1$.
In this case, $d_k$ and nearby bits can be flipped to another $d_k'$ by perturbation because the eigenvalue difference between these states is $\sim \max\{J_{i,k}, J_{k,i+c-1}\}$, which is exponentially small.
Note that $d_l\:(d_l=3,4)$ corresponds the diagonal elements $\ket{+1}\bra{+1}$ or $\ket{-1}\bra{-1}$ in the original density matrix before vectorization.
In this sense, some eigenstates exist for which the diagonal degrees of freedom form large delocalized clusters.

On the other hand, 
the states  $|\phi^1_l)$  and 
$|\phi^2_l)$, which correspond to the off-diagonal degrees of freedom in the density matrix, are typically stable under (first-order) perturbation.
Here, we explain some of the prototypical examples for which the transition processes are suppressed by strong disorder.
For the single flip from $o_l\:(o_l=1,2)$ [$o$ stands for off-diagonal degree of freedom] to $d_l\:(d_l=3,4)$, while the transition matrix element is $\sim \gamma(Z_l^{x/y})\sim g\gamma/\tilde{h}_l$, the eigenvalue difference 
becomes $\sim |\pm 2i\tilde{h}_l+2(Z_l^z)^2\gamma|$, so the ratio becomes $\sim g\gamma/h^2\ll 1$.
According to perturbation theory, this indicates that the transition process is suppressed due to the large energy mismatch.
For the single flip from $o_l\:(o_l=1,2)$ to $o'_l\:(o'_l=2,1)$ with  the transition matrix element $\sim \gamma{(Z_l^{x/y})}^2 \sim g^2\gamma/\tilde{h}_l^2$, the eigenvalue difference 
becomes $\sim |4i\tilde{h}_l|$, and therefore the ratio again becomes small, of the order of $g^2\gamma/h^3\ll 1$.
For another example of the double flip from $|\cdots o_l\cdots d_m \cdots)$ to 
$|\cdots d_l\cdots o_m \cdots)$ with the transition matrix element  $\mr{O}(g^2\gamma/{h}^2)$ and exponentially small with respect to $|l-m|$, the eigenvalue difference 
becomes $\sim |\pm 2i(\tilde{h}_l-\tilde{h}_m)+2\{(Z_l^z)^2\gamma-(Z_m^z)^2\gamma\}|$, and therefore the ratio is again small unless $\tilde{h}_l-\tilde{h}_m$ is accidentally small.

In conclusion, for sufficiently strong disorder, the off-diagonal degrees of freedom can be localized.
The decay rate is then stabilized at 
\aln{
\mr{Im}[\lambda_\alpha]= \sum_l2\gamma(\delta_{\alpha_l,1}+\delta_{\alpha_l,2})+\mr{O}\lrs{\lrs{\frac{g}{h}}^2}.
}
If there are $q$ localized off-diagonal degrees of freedom, the decay rate is $\sim 2\gamma q$, as discussed in the main text.

\begin{figure}
\begin{center}
\includegraphics[width=\linewidth]{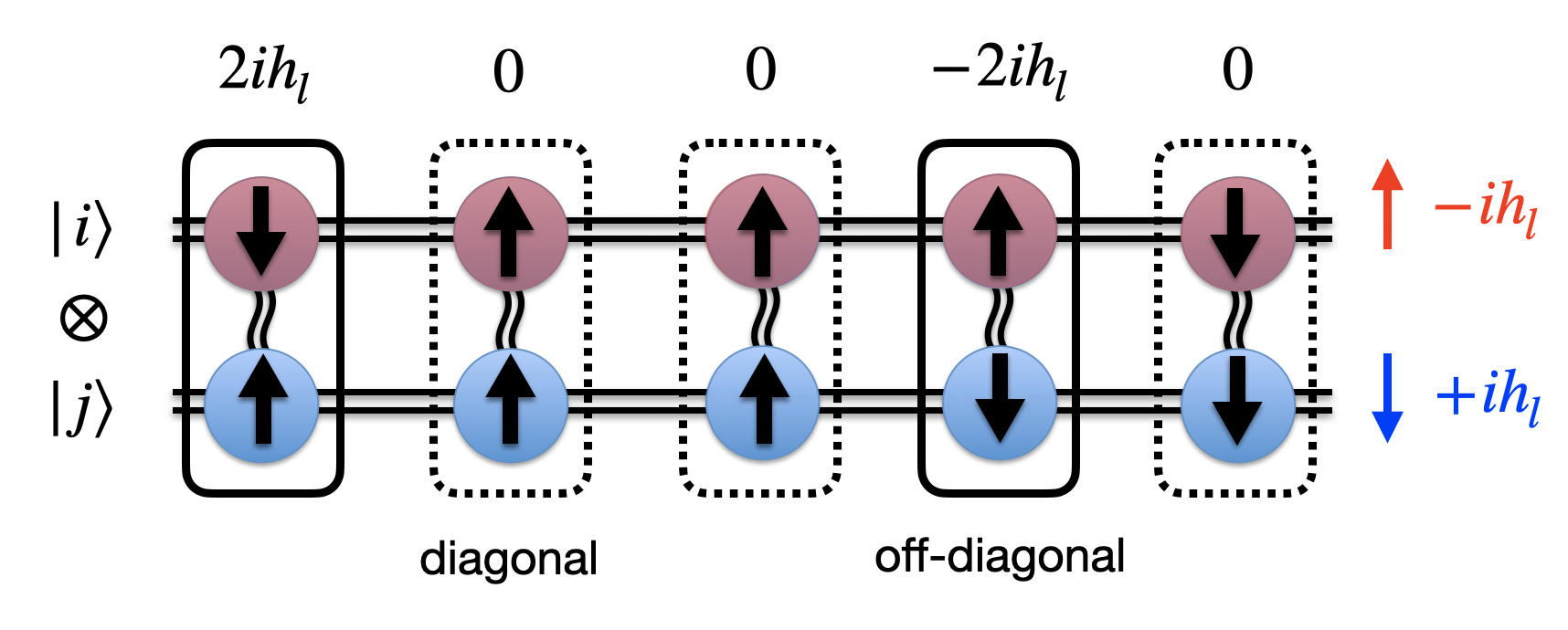}
\end{center}
\caption{
Schematic illustration of the Lindbladian in the matrix representation. Black arrows show quasi-local bits in the $\hat{\tau}^z$ basis. For diagonal degrees of freedom (surrounded by dotted lines), the net disorder vanishes and delocalization occurs.
On the other hand, off-diagonal degrees of freedom (surrounded by solid lines) undergo finite net disorder $\mp 2ih_l$, and localization occurs.
}
\label{ladder}
\end{figure}

The localization/delocalization of off-diagonal/diagonal degrees of freedom is intuitively understood from the matrix representation of the Lindbladian.
As shown in Fig.~\ref{ladder}, the Lindbladian includes a term $-ih_l(\hat{\tau}_l^z\otimes\hat{\mbb{{I}}}-\hat{\mbb{{I}}}\otimes\hat{\tau}_l^z)$. 
For off-diagonal degrees of freedom $|\phi^1)$ and $|\phi^2)$, they experience  net disorder $\mp 2ih_l$ and thus localize.
On the other hand, diagonal degrees of freedom $|\phi^3)$ and $|\phi^4)$ do not undergo disorder and thus delocalize.

To discuss the operator-space entanglement entropy (OSEE), 
we notice that eigenstates  in the middle of the spectrum have $\sim (1-\alpha)L\:\:(\alpha <1)$ localized off-diagonal degrees of freedom.
Thus, the number of the remaining diagonal degrees of freedom in those states is $\sim\alpha L$.
Since the localized degrees of freedom do not contribute to the OSEE (except for the exponential tail coming from the difference between $\hat{\tau}_l$ and the physical bits $\hat{\sigma}_l$), the half-chain OSEE for strong disorder is typically bounded from above by $\sim \alpha L\log 2$.
This is generally smaller than the half-chain OSEE for the weak-disorder case, where the RMT prediction leads to $\sim L\log 2$.
We also note that, if the clusters of the diagonal degrees of freedom are small and kept far apart from one another by many localized off-diagonal clusters, they cannot create much entanglement.
In that case, the OSEE can be further reduced from $\sim \alpha L\log 2$ and satisfy the area law $\sim \mr{O}(L^0)$.
Thus, depending on the number and the position of the localized off-diagonal degrees of freedom, the OSEE can satisfy either the volume law $\sim \alpha L\log 2$ or the area law $\sim \mr{O}(L^0)$.
In the middle of the spectrum, we expect  many localized off-diagonal clusters and the area-law behavior of the OSEE.
This is consistent with the numerical simulation in the main text, but
we leave it for a future problem to investigate the precise evaluation for larger system sizes.
In any case, at the transition between these two phases, there appears a large fluctuation of the OSEE for different eigenstates, leading to the observed peak in the main text.

We also note that, even when a large delocalized cluster of diagonal degrees of freedom exists, it may not have the power to delocalize the localized off-diagonal degrees of freedom, which is distinct from the situation where the large bath can delocalize the localized bits in the Hermitian disordered systems~\cite{Roeck17}.
To see this, 
let us consider a simple situation where a single localized off-diagonal bit, $|\phi_1^1)$ (or $|\phi_1^2)$), is coupled to a delocalized cluster of diagonal degrees of freedom $|D)$.
For the resonant transition from $|\phi_1^1)\otimes |D)$ to $|\phi_1^\alpha)\otimes |D')\:(\alpha=2,3,4)$ to occur under some small perturbation, where $|D')$ is another state for the delocalized cluster, the difference between the imaginary parts of the energies of  $D$ and $D'$ should be of the order of $h$.
However, since even large delocalized clusters have almost zero imaginary parts of energies, which correspond to the eigenvalues in the original language, this resonance condition is typically not  satisfied.

\subsection{Case of damping}
We briefly discuss the damping case.
To do this, we decompose the Hamiltonian as $\hat{H}=\hat{H}_{NI}+\hat{H}_I$, where $\hat{H}_{NI}=\sum_l\tilde{h}_l\hat{\tau}_l^z$ is the non-interacting term of $\hat{\tau}_l^z$ and $\hat{H}_I$ is the interaction term.
Then, we have 
\aln{
\tilde{\mc{L}}=-i(\hat{H}_{NI}\otimes\mbb{\hat{I}}-\hat{\mbb{I}}\otimes\hat{H}_{NI}^T)+\sum_l\frac{{Z_l^z}^2\gamma}{4}(2\hat{\tau}_l^-\otimes\hat{\tau}_l^--\hat{\tau}_l^+\hat{\tau}_l^-\otimes\hat{\mbb{I}}-\hat{\mbb{I}}\otimes\hat{\tau}_l^+\hat{\tau}_l^-)+\tilde{\mc{L}}_I+\tilde{\mc{L}}_P,
}
where $\tilde{\mc{L}}_I=-i(\hat{H}_{I}\otimes\mbb{\hat{I}}-\hat{\mbb{I}}\otimes\hat{H}_{I}^T)$.
We first consider $\tilde{\mc{L}}_I+\tilde{\mc{L}}_P$ as a perturbation.

The eigenstates of $\mc{L}-\mc{L}_I-\mc{L}_P$ can be written as 
 \aln{
|{R}_\alpha)=|\eta^{\alpha_1}_1)|\eta^{\alpha_2}_2)\cdots |\eta^{\alpha_L}_L)
 }
and
\aln{
|{L}_\alpha)=|\chi^{\alpha_1}_1)|\chi^{\alpha_2}_2)\cdots |\chi^{\alpha_L}_L),
 }
 where
 $|\eta^1)=|\chi^1)=|+1,-1)$,  $|\eta^2)=|\chi^2)=|-1,+1)$,  $|\eta^3)=\frac{|+1,+1)-|-1,-1)}{\sqrt{2}}$, 
 $|\chi^3)=|+1,+1)$, $|\eta^4)=|-1,-1)$, and $|\chi^4)=\frac{|+1,+1)+|-1,-1)}{\sqrt{2}}$.
The corresponding eigenvalue is given by
\aln{\label{eigdam}
-i\sum_l2\tilde{h}_l(\delta_{\alpha_l,1}-\delta_{\alpha_l,2})-i(\text{terms including } J_{lm}, \cdots)-\sum_l{Z_l^z}^2\gamma(\delta_{\alpha_l,1}+\delta_{\alpha_l,2}+2\delta_{\alpha_l,3}).
}

Even if we add $\tilde{\mc{L}}_I+\tilde{\mc{L}}_P$, the states $|\eta^1_l)$ and $|\eta^2_l)$ are robust under the first-order perturbation.
Indeed, just as the dephasing case, changing these states to other states is typically energetically costly due to strong disorder.
This leads to the localization of these off-diagonal degrees of freedom and the breakdown of non-Hermitian random-matrix-type universality.

We also note that, 
the longest-lived modes involving $q$ off-diagonal degrees of freedom
have stabilized decay rates $\sim \gamma q(1+\mr{O}((g/h)^2))$.
To see this, we next consider only $\tilde{\mc{L}}_P$ as a perturbation as in the dephasing case.
Without  perturbation, we notice that   states such as
\aln{\label{llm}
|\eta^{4}_1)|\eta^{4}_2)\cdots |\eta^{4}_{L-q})|\eta^{1\text{ or }2}_{L-q+1})\cdots |\eta^{1\text{ or }2}_L)
}
are eigenstates of the unperturbed Lindbladian in the matrix representation with the eigenvalue $\gamma q(1+\mr{O}((g/h)^2))$.
Note that the eigenstates involving $|\eta^3)$ have a faster decay rate, so the states represented in Eq.~\eqref{llm} are those having the longest lifetime with $q$ off-diagonal degrees of freedom.
When we introduce the perturbation $\tilde{\mc{L}}_P$, the leading order of the change of the eigenvalue is $\mr{O}(\gamma(g/h)^2))$, and the eigenvalue is still expressed as $\gamma q(1+\mr{O}((g/h)^2))$.

\section{Effects of dissipation on the Lindbladian MBL transition point}
Here, we discuss how the Lindbladian MBL transition depends on the dissipation strength $\gamma$.

\begin{figure}
\begin{center}
\includegraphics[width=\linewidth]{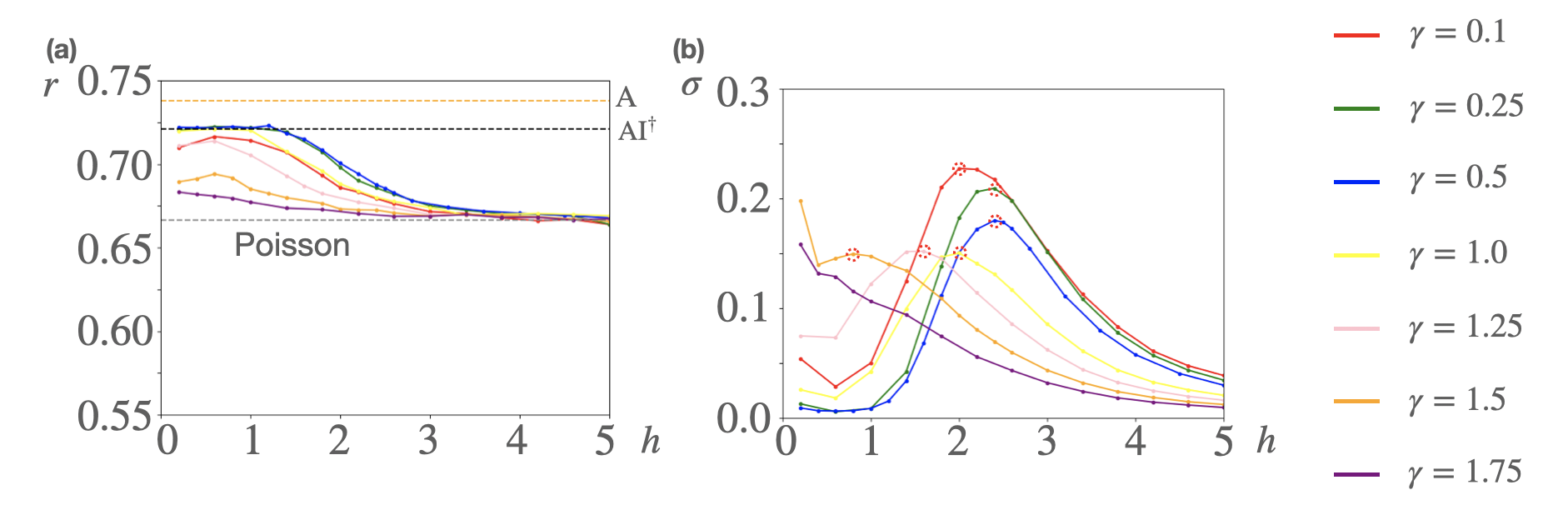}
\end{center}
\caption{
Complex-spacing ratio $r$ (a) and the variance $\sigma$ of the OSEE (b) as a function of $h$ for different values of $\gamma$.
We show the results for a dephasing type of dissipation.
We find that $r$ changes from the universal value of non-Hermitian RMT (class AI$^\dag$) to the Poisson value as $h$ increases, unless $\gamma$ is too small or too large.
On the other hand, the critical value of $h$ depends on $\gamma$.
For small $\gamma\:(\lesssim 0.5)$, an increase in $\gamma$ increases the critical value of the transition, which is read out from the peak of $\sigma$ (dotted circles in panel (b)).
On the other hand, a further increase in $\gamma$ decreases the critical value.
When $\gamma$ is large enough (e.g., $\gamma=1.75$), the peak no longer appears, which indicates that the system belongs to the localized phase even for weak $h$.
We use the Lindbladian with $J=1, g=-0.9, L =6$ and eigenvalues/eigenstates in the middile of the supectrum~\cite{Mid_foot}. 
The number of samples used for the simulation is $800$.
}
\label{gammavary}
\end{figure}

\begin{figure}
\begin{center}
\includegraphics[width=\linewidth]{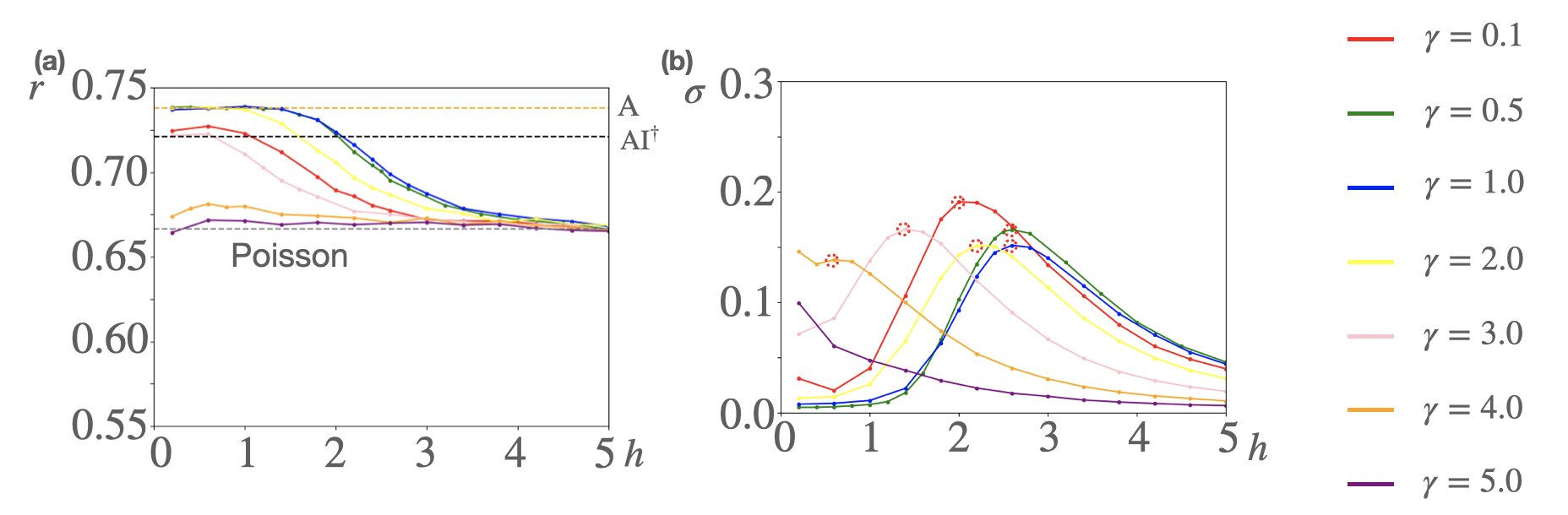}
\end{center}
\caption{
Complex-spacing ratio $r$ (a) and the variance $\sigma$ of the OSEE (b) as a function of $h$ for different values of $\gamma$.
We show the results for a damping type of dissipation.
We find that $r$ changes from the universal value of non-Hermitian RMT (class A) to the Poisson value as $h$ increases, unless $\gamma$ is too small or too large.
On the other hand, the critical value of $h$ depends on $\gamma$.
For small $\gamma\:(\lesssim 1)$, increasing $\gamma$ increases the critical value of the transition, which is read out from the peak of $\sigma$ (dotted circles in panel (b)).
On the other hand, a further increase in $\gamma$ decreases the critical value.
When $\gamma$ is strong enough (e.g., $\gamma=5.0$), the peak no longer appears, which indicates that the system belongs to the localized phase even for weak $h$.
We use the Lindbladian with $J=1, g=-0.9, L =6$ and eigenvalues/eigenstates in the middile of the supectrum~\cite{Mid_foot}. 
The number of samples used for the simulation is $800$.}
\label{gammavarydam}
\end{figure}

\begin{figure}
\begin{center}
\includegraphics[width=\linewidth]{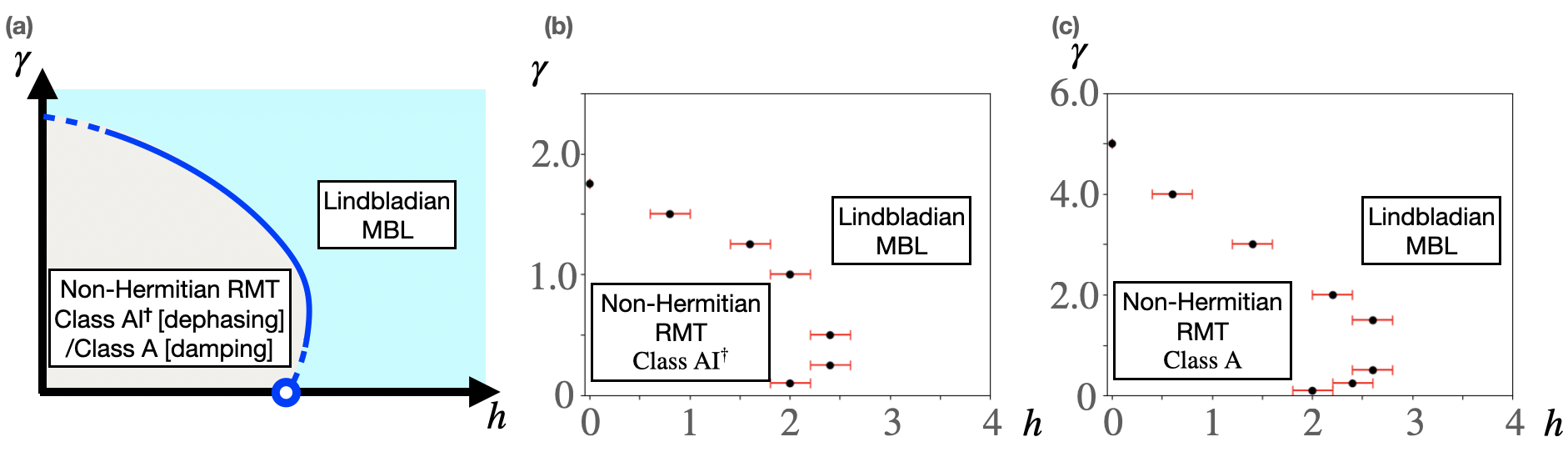}
\end{center}
\caption{
(a) Schematic illustration of the phase diagram of dissipative disordered systems.
There are two phases: a phase characterized by the non-Hermitian RMT and that characterized by the Lindbladian MBL.
(b) Phase diagram for the dephasing case. Each transition point is determined from the maximum peak of the variance of the OSEE for $L=6$, as shown in Fig.~\ref{gammavary}(b).
(c) Phase diagram for the damping case. Each transition point is determined from the maximum peak of the variance of the OSEE for $L=6$, as shown in Fig.~\ref{gammavarydam}(b).
While the phase boundaries are quantitatively different for (b) and (c), both of them are qualitatively described by the non-monotonic phase boundary given in (a).
}
\label{phase}
\end{figure}

\subsection{Phase diagram}
We first discuss the possible phase diagram of our system.
Figures~\ref{gammavary} and~\ref{gammavarydam} show the $h$-dependence of the complex-spacing ratio $r$ and that of the variance  $\sigma$ of the OSEE for different values of $\gamma$ for the cases of dephasing and damping, respectively.
We find that $r$ changes from the universal value of non-Hermitian RMT to the Poisson value as $h$ increases, as long as $\gamma$ is not too small.
On the other hand, the critical value depends on $\gamma$.
For small $\gamma\:(\lesssim 0.5)$, increasing $\gamma$ increases the critical value of the transition, which is read out from the peak of $\sigma$.
On the other hand, a further increase in $\gamma$ decreases the critical value.
If $\gamma$ is strong enough, the peak no longer appears, which indicates that the system belongs to the localized phase even for weak $h$.

From these results, we can identify the phase diagram of our system with each type of dissipation, as shown in Fig.~\ref{phase}.
Two remarks are in order here.
First, the transition points $h_c(\gamma)$ exhibit a non-monotonic behavior.
Second, a clear quantitative difference exists for the transition points of dephasing-type and damping-type cases.
These two points indicate a nontrivial interplay between dissipation and disorder in this system.

Note that the phase diagrams in Fig.~\ref{phase} are obtained for relatively small system sizes ($L=6$). When we increase the system size, the exact values of the transition points may change. However, we conjecture that the qualitative feature of the phase diagram remains the same for larger system sizes.

\subsection{Relation to the many-body localization without dissipation}
Here, we discuss the relation between the Lindbladian MBL and the conventional MBL that occurs in our Ising model $\hat{H}$ without dissipation (i.e., $\gamma=0$).
While this short-range interacting model is believed to exhibit MBL for strong disorder~\cite{Imbrie16D,Imbrie16O}, not much is investigated about the transition behavior between delocalized and localized phases (see, e.g., Ref.~\cite{PhysRevB.102.104302} for an exception).

\begin{figure}
\begin{center}
\includegraphics[width=\linewidth]{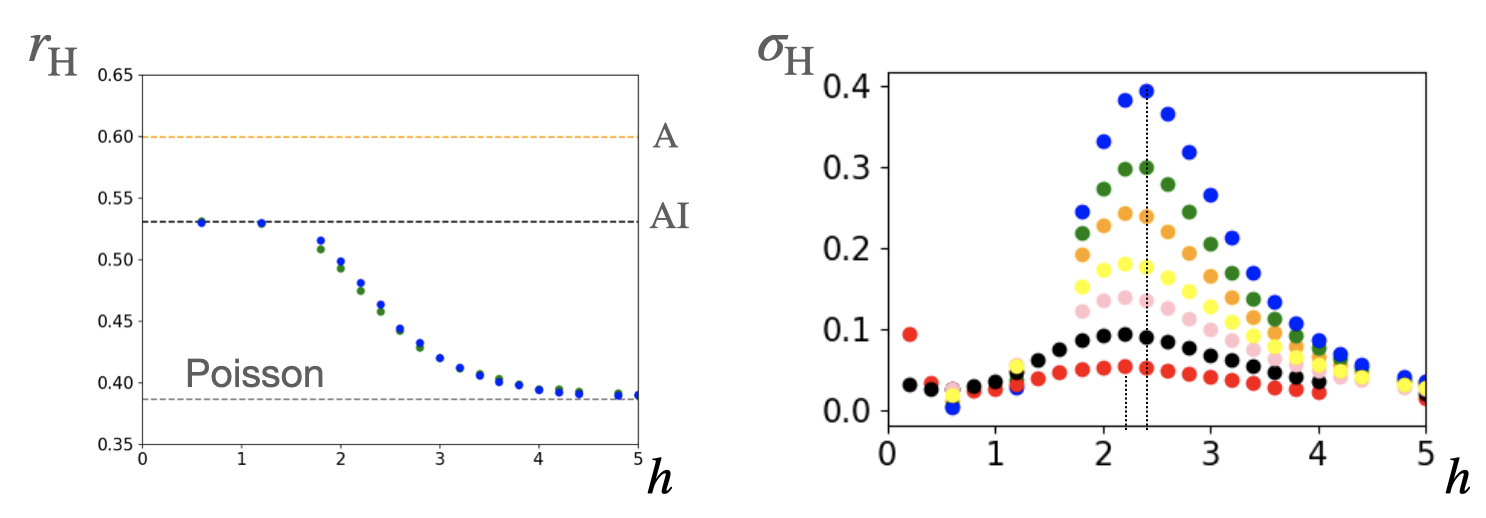}
\end{center}
\caption{
Many-body localization transition in the Hermitian quantum Ising model.
(a) The spacing ratio $r_\mr{H}$ of (real) energy eigenvalues of $\hat{H}$.
It changes from $r_\mr{H,AI}\simeq 0.5307$ (the RMT prediction corresponding to Hermitian class AI) to  $r_\mr{H, Po}\simeq 0.3863$ (the Poisson value), as $h$ is increased.
We use $L=11$ (green) and $12$ (blue).
(b)
The variance of the entanglement entropy of energy eigenstates.
The value of the peak becomes larger as we increase the system size from $L=6$ (red) to $12$ (blue).
From the peak for $L=12$, we can estimate the MBL transition point as $h_{\mr{H},c}= 2.4\pm 0.2$.
On the other hand, the peak position for $L=6$ is slightly shifted to $ 2.2\pm 0.2$.
For both panels, we use the Hamiltonian with $J=1$ and $g=-0.9$. 
We use the eigenstates in the middle of the spectrum (6.25\%).
The number of samples used for the simulation is $10000\: (L=6,7,8,9,10)$ and $800\: (L=11,12)$.
}
\label{herm_MBL}
\end{figure}

We numerically investigate the MBL transition for $\hat{H}$ with the same parameters in the Hamiltonian as those used in the main text.
For this purpose, we first calculate the spacing ratio of (real) energy eigenvalues of $\hat{H}$~\cite{PhysRevLett.110.084101}, defined by $r_\mr{H}=\mbb{E}_a\lrl{\frac{\min{(E_{a+1}-E_{a},E_a-E_{a-1})}}{\max{({E_{a+1}-E_{a},E_a-E_{a-1}})}}}$.
As shown in Fig.~\ref{herm_MBL}(a), it changes from $r_\mr{H,AI}\simeq 0.5307$ (the RMT prediction corresponding to Hermitian class AI) to  $r_\mr{H, Po}\simeq 0.3863$ (the Poisson value), as $h$ is increased.

We next calculate the eigenstate variance $\sigma_\mr{H}$ of the  entanglement entropy $S_{\mr{H},a}(\ket{E_a})=-\Tr[\hat{\rho}_a^l\ln\hat{\rho}_a^l]$ of the energy eigenstates $\ket{E_a}$, where $\hat{\rho}_a^l$ is the reduced density matrix of $\ket{E_a}$ to a region $[1,l]\:(l=\lfloor L/2\rfloor)$.
As shown in Fig.~\ref{herm_MBL}(b), $\sigma_\mr{H}$ has a peak when $h$ is varied, from which we can read out the MBL transition point $h_{\mr{H},c}$~\cite{Kjall14}.
We can estimate that the transition point as $h_{\mr{H},c}= 2.4\pm 0.2$ from the results up to $L=12$.

A natural question is whether the Lindbladian MBL transition point $h_c(\gamma)$ coincides with the Hermitian MBL transition point $h_{\mr{H},c}$ in the dissipationless limit $\gamma\ra 0$.
From Fig.~\ref{phase}, we can see that $h_c(\gamma\ra 0)\leq 2.0\pm 0.2$.
This is different from the above estimate of $h_{\mr{H},c}=2.4\pm 0.2$ obtained from $L=12$.
On the other hand, since $h_c(\gamma)$ is numerically estimated from $L=6$, it is more natural to compare the result with $h_{\mr{H},c}$ obtained from the data for $L=6$~\footnote{Note that the true transition point should be determined from the limit of $L\ra \infty$. We here compare the estimation of the transition points for the same system size to take account of the degree of the finite-size effects appropriately.}.
From Fig.~\ref{herm_MBL}(b), this corresponds to $h_{\mr{H},c}=2.2\pm 0.2$, which is consistent with the value of $h_c(\gamma)$ with small $\gamma$.

To discuss behavior in the thermodynamic limit, let us assume that the OSEE of the Lindbladian eigenstates is continuous as a function of $\gamma$.
If that is the case, the OSEE can be obtained as
\aln{
\lim_{\gamma\ra 0} S_\alpha(\gamma)= S_\alpha(\gamma=0) = S_{\mr{H},a}+S_{\mr{H},b},
}
where we have assumed $\hat{R}_\alpha(\gamma=0)=\ket{E_a}\bra{E_b}$ with $\lambda_\alpha=-i(E_a-E_b)$.
For simplicity, we focus on the statistical behavior of $ S_\alpha(\gamma=0)$ for eigenstates satisfying $-i\lambda_\alpha\sim \omega\in\mbb{R}$.
The total number of such eigenstates becomes 
$\int dE e^{\mc{S}(E)+\mc{S}(E+\omega)}$, where $\mc{S}(E)$ is the thermodynamic entropy of the system.
While this number contains multiple energy scales, for large system sizes, the contribution coming from $E=\mc{E}$ that satisfies 
\aln{
\fracd{\mc{S}}{E}(\mc{E})+\fracd{\mc{S}}{E}(\mc{E}+\omega)=0
}
dominates within
the saddle-point approximation.
Then, 
$S_\alpha(\gamma=0)$ for almost all $\alpha$ behaves as
$S_{\mr{H},a}+S_{\mr{H},b}$ with $E_a\simeq \mc{E}$ and $E_b\simeq \mc{E}+\omega$.
For example, the average of $S_\alpha(\gamma=0)$ over eigenstates satisfies $S(\lambda=i\omega)\simeq S_\mr{H}(\mc{E})+S_\mr{H}(\mc{E}+\omega)$, and similarly, the variance satisfies
$\sigma(\lambda=i\omega)\simeq \sigma_\mr{H}(\mc{E})+\sigma_\mr{H}(\mc{E}+\omega)$, provided that there is no correlation between fluctuations of $S_{\mr{H},a}$ and $S_{\mr{H},b}$.

From the above discussion, we can relate $h_c(\gamma\ra 0)$ to $h_{\mr{H},c}$ if we further make either of the following two assumptions: (i) $\omega=\mr{o}(L)$, or (ii) there is no mobility edge, i.e., the MBL transition occurs for all energy scales simultaneously.
If we assume (i), we have $S(\lambda=i\omega)\simeq 2S_\mr{H}(\mc{E})$ and $\sigma(\lambda=i\omega)\simeq 2\sigma_\mr{H}(\mc{E})$. 
We also have $\fracd{\mc{S}}{E}(\mc{E})=0$, indicating that $\mc{E}$ corresponds to the energy whose entropy is maximal.
Then, the MBL transition (i.e., the transition that $S_\mr{H}$ changes to the area law and that $\sigma_\mr{H}$ has a peak) at $\mc{E}$ leads to the Lindbladian MBL.
In other words, $h_{\mr{H},c}$ at infinite temperature and $h_c(\gamma\ra 0)$ with small $\omega=\mr{o}(L)$ coincide in the thermodynamic limit.
Instead, if we assume (ii), whose possibility is discussed in Ref.~\cite{PhysRevB.93.014203}, the MBL transition point $h_{\mr{H},c}$ does not depend on $E$.
Thus, in crossing $h_{\mr{H},c}$, $S_\mr{H}(\mc{E})$ and $S_\mr{H}(\mc{E}+\omega)$ change to the area law, and 
$\sigma_\mr{H}(\mc{E})$ and $\sigma_\mr{H}(\mc{E}+\omega)$ have a peak.
This directly leads to the change from the volume to the area laws for $S(\lambda)$ and the emergence of a peak for $\sigma(\lambda)$.
In other words, $h_{\mr{H},c}$ and $h_c(\gamma\ra 0)$  coincide in the thermodynamic limit.
It is a future problem to numerically investigate if either (or both) of the above two scenarios holds for large system sizes.

\bibliographystyle{apsrev4-1}
\bibliography{derh2.bib}